\documentclass[aps,prb,reprint,superscriptaddress,twocolumn,showpacs,notitlepage]{revtex4-1}
\setcitestyle{numbers,square}

\usepackage[T1]{fontenc}
\usepackage[utf8]{inputenc}
\usepackage{textcomp}
\usepackage{color}
\usepackage{comment}
\usepackage{braket}
\usepackage{amsmath}
\usepackage{array,mathtools,amssymb,booktabs}
\usepackage{multirow}
\usepackage{verbatim}
\usepackage{textgreek}

\usepackage{gensymb}
\usepackage{csquotes} 

\usepackage{lipsum}

\usepackage{xcolor}
\definecolor{green2}{RGB}{50,160,50}

\usepackage{hyperref}
\hypersetup{
  colorlinks= True,
  linkcolor	=black,
  citecolor	=blue,
  urlcolor	=blue
}

\newcommand{\ang}		{\text{\normalfont\AA}}

\begin{document}
\title{
Sensitivity of the electronic and magnetic structures of cuprate superconductors to density functional approximations} 

\author{Kanun Pokharel}
\email[]{kpokhare@tulane.edu}
\affiliation{Department of Physics and Engineering Physics, Tulane University, Louisiana 70118 New Orleans, USA}

\author{Christopher Lane}
\email{laneca@lanl.gov}
\affiliation{Theoretical Division, Los Alamos National Laboratory, Los Alamos, New Mexico 87545, USA}
\affiliation{Center for Integrated Nanotechnologies, Los Alamos National Laboratory, Los Alamos, New Mexico 87545, USA}

\author{James W. Furness}
\affiliation{Department of Physics and Engineering Physics, Tulane University, Louisiana 70118 New Orleans, USA}

\author{Ruiqi Zhang}
\affiliation{Department of Physics and Engineering Physics, Tulane University, Louisiana 70118 New Orleans, USA}

\author{Jinliang Ning}
\affiliation{Department of Physics and Engineering Physics, Tulane University, Louisiana 70118 New Orleans, USA}

\author{Bernardo Barbiellini}%
\affiliation{LUT University, P.O. Box 20, FI-53851, Lappeenranta, Finland}
\affiliation{Physics Department, Northeastern University, Boston, Massachusetts 02115, USA}

\author{Robert S. Markiewicz}
\affiliation{Physics Department, Northeastern University, Boston, Massachusetts 02115, USA}

\author{Yubo Zhang}
\affiliation{Department of Physics and Engineering Physics, Tulane University, Louisiana 70118 New Orleans, USA}

\author{Arun Bansil}
\email[]{ar.bansil@neu.edu}
\affiliation{Physics Department, Northeastern University, Boston, Massachusetts 02115, USA}

\author{Jianwei Sun}
\email[]{jsun@tulane.edu}
\affiliation{Department of Physics and Engineering Physics, Tulane University, Louisiana 70118 New Orleans, USA}%

\date{\today}

\begin{abstract}
We discuss the crystal, electronic, and magnetic structures of $\mathrm{La_{2-x}Sr_{x}CuO_{4}}$ (LSCO) for $x=0.0$ and $x=0.25$ employing 13 density functional approximations, representing the local, semi-local, and hybrid exchange-correlation approximations within the Perdew—Schmidt hierarchy. The meta-generalized gradient approximation (meta-GGA) class of functionals is found to perform well in capturing the key properties of LSCO, a prototypical high-temperature cuprate superconductor. In contrast, the local-spin-density approximation, GGA, and the hybrid density functional fail to capture the metal-insulator transition under doping. 
  
\end{abstract}

\maketitle

\section*{Introduction}
Ever since the discovery of cuprate superconductivity in 1986 by Bednorz and M\"uller\cite{bednorz1986possible}, the anomalous behavior of the pristine as well as the doped cuprate has eluded theoretical explanation  and still remains an unsolved problem in condensed matter physics. $\mathrm{La_{2}CuO_{4}}$ (LCO), in particular, has been a significant challenge to describe within a  coherent theoretical framework. The Hohenberg-Kohn-Sham density functional theory (DFT)\cite{hohenberg1964inhomogeneous,kohn1965self} with some classes of popular exchange-correlation (XC) approximations fails spectacularly to capture the insulating antiferromagnetic ground state of LCO, let alone the metal insulator transition (MIT) under doping. \cite{pickett1989electronic}. Specifically, the local spin-density approximation (LSDA) XC functional incorrectly predicts the parent compound to be a metal, yielding a vastly underestimated value for the copper magnetic moment of 0.1${\mathrm{\mu}}_B$ \cite{mattheiss1987electronic,perdew1981self} compared to the experimental value of 0.60$\pm$0.05 $\mu_{B}$\cite{tranquada2007handbook}. The  PBE generalized gradient approximation (GGA) \cite{perdew1996generalized} still predicts LCO to be a metal with a slightly improved magnetic moment of 0.2$ {\mathrm{\mu}}_B$ \cite{singh1991gradient}. The Becke-3-Lee-Yang-Parr (B3LYP)\cite{becke1988density,lee1988development,becke1993new,stephens1994ab} hybrid functional correctly explains the AFM ground state in LCO but fails to capture the MIT upon doping \cite{perry2002ab}. 
These failures have led to the (incorrect) belief that DFT is fundamentally incapable of capturing the physics of the cuprates and other correlated materials. Therefore, “beyond DFT” methodologies, such as the quantum Monte Carlo methods \cite{wagner2014effect}, DFT+U \cite{czyzyk1994local,pesant2011dft+}, and dynamical mean-field theory (DMFT)\cite{kotliar2006electronic,held2006realistic,park2008cluster} have been introduced to handle strong electron correlation effects. These approaches have been useful for understanding the physics of the cuprates, although they typically introduce {\it ad hoc} parameters, such as the Hubbard U, to tune the correlation strength, which limits their predictive power.

Recent progress in constructing advanced density functional approximations (DFA) provides a viable new pathway for addressing the electronic structures of correlated materials. In particular, the strongly-constrained and appropriately-normed (SCAN) meta-GGA \cite{sun2015strongly} , which obeys all 17 known constraints applicable to a meta-GGA functional, has been shown to accurately predict many key properties of the pristine and doped $\mathrm{La_{2}CuO_{4}}$ and $\mathrm{YBa_{2}Cu_{3}O_{6}}$ \cite{furness2018accurate,lane2018antiferromagnetic,zhang2020competing}. In LCO, SCAN correctly captures the size of optical band gap, the magnitude and the orientation of the copper magnetic moment, and the magnetic form factor in comparison with the corresponding experimental results \cite{lane2018antiferromagnetic}. In near-optimally doped $\mathrm{YBa_{2}Cu_{3}O_{7}}$, 26 competing uniform and stripe phases are identified \cite{zhang2020competing}. In this case, the treatment of charge, spin, and lattice degrees of freedom on the same footing is crucial in stabilizing the stripe phases without invoking any free parameters.  Furthermore, SCAN has been applied to  the $\mathrm{Sr_{2}IrO_{4}}$ parent compound yielding the subtle balance between electron correlations and strong spin-orbit coupling in excellent agreement with experiment\cite{lane2020first}. 

SCAN’s success in the copper and iridium oxides is a significant achievement for the DFT and suggests capability for treating a wider class of correlated materials. SCAN, however, is well-known to yield overly large saturation magnetization in elemental metals (e.g., Fe and Ni) \cite{isaacs2018performance, ekholm2018assessing,fu2018applicability} and has the problem of numerical instabilities \cite{yang2016more,furness2019enhancing,bartok2019regularized,furness2020accurate}, which may limit its applicability. A number of natural questions therefore arise: Is SCAN a  unique XC density functional that is able to correctly capture a variety of properties of the cuprates or do other meta-GGAs perform similarly well? How do hybrid XC functionals perform in comparison? Answers to these questions are important for benchmarking the performance of SCAN and related  DFAs, and for opening a pathway to their more extensive use. 

With this motivation, this paper compares the accuracy of 13  DFAs. In particular, we assess the efficacy of LSDA \cite{jones1989density,perdew1992accurate}, PBE\cite{perdew1996generalized}, SCAN\cite{sun2015strongly}, SCAN-L\cite{mejia2017deorbitalization}, rSCAN\cite{bartok2019regularized}, $\mathrm{r}^{2}$SCAN\cite{furness2020accurate}, $\mathrm{r}^{2}$SCAN-L\cite{mejia2020meta}, TPSS\cite{tao2003climbing}, revTPSS\cite{perdew2009workhorse}, MS0\cite{sun2012communication}, MS2\cite{sun2013semilocal}, M06L\cite{zhao2006new}, and HSE06\cite{heyd2006j,heyd2005energy,heyd2004j,peralta2006spin}
with respect to their predictions for crystal, electronic, and magnetic structures of the pristine and doped prototypical high-temperature superconductors $\mathrm{La_{2-x}Sr_{x}CuO_{4}}$. Various XC density functionals employed span the levels of the Perdew—Schmidt hierarchy\cite{perdew2001density}, allowing us to evaluate the performance of each functional class for the description of correlated condensed matter systems.

\section*{RESULTS and DISCUSSION}

\subsection*{Methodology}
The theoretical foundation of DFT was laid by Hohenberg and Kohn  \cite{hohenberg1964inhomogeneous} when they considered the {\it electron density} rather than the wave function as the fundamental object for addressing the many-body problem. This concept was later extended by Kohn and Sham (KS)  by replacing the complicated many-electron problem with an auxiliary non-interacting system that leads to the one-electron Schr\"odinger like equations \cite{kohn1965self}, providing a practical approach to solve for the ground-state electron density and ground-state energy of the many body system. The beauty of KS approach is that it explicitly separates the non-interacting kinetic energy and the long-range Hartree energy, which describes the classical electrostatic repulsion between electrons, from  the  remaining  exchange-correlation  energy. An exact solution for the ground-state total energy and electron density is then obtained, in principle, although in practice, the exchange-correlation energy must be approximated. 

The total energy of the many-body electron system, within the Kohn-Sham DFT framework can be written as:
\begin{equation} \label{eq:dft1}
\mathrm{{E}} = \mathrm{T_{s}}+\mathrm{E_{ext}}+\mathrm{E_{H}}+\mathrm{E_{xc}},
\end{equation}
where $\mathrm{T_{s}}$ is the non-interacting kinetic energy, $\mathrm{E_{ext}}$ is the external potential energy, $\mathrm{E_{H}}$ is the Hartree energy, and $\mathrm{E_{xc}}$ contains the remaining energy contributed by the many-body XC effects. The first three terms in Eq.~\ref{eq:dft1} can be obtained exactly while the last term has to be approximated. Various approximations for $\mathrm{E_{xc}}$ can be arranged on the rungs of the Perdew-Schmidt hierarchy
\cite{perdew2001density}. The lowest rung of the hierarchy is LSDA that is based on local electron densities $n_{\sigma}$. On the next rung, the GGA class adds density gradients $\nabla n_{\sigma}$ to LSDA. This is followed by meta-GGAs that come in two flavors, adding either the non-interacting kinetic energy density $\tau_{\sigma}$ or a Laplacian dependence $\nabla^{2} n_{\sigma}$ or both in comparison with GGAs. Here $\sigma$ denotes the two spin channels. The XC energy for meta-GGAs is defined as

\begin{align}
&\mathrm{E}^{\mathrm{MGGA}}_{\mathrm{xc}}[\mathrm{n}_{\uparrow},\mathrm{n}_{\downarrow}] =\\
&\int \mathrm{d^3}\mathrm{\mathbf{r}}\mathrm{n}(\mathrm{\mathbf{r}})\mathrm{\varepsilon}^{\mathrm{unif}}_{\mathrm{x}}(\mathrm{n}) \mathrm{F}_{\mathrm{xc}}(\mathrm{n}_{\uparrow},\mathrm{n}_{\downarrow},\nabla{\mathrm{n}}_{\uparrow},\nabla{\mathrm{n}}_{\downarrow},\nabla^2\mathrm{n}_{\uparrow},\nabla^2\mathrm{n}_{\downarrow},\tau_{\uparrow},\tau_{\downarrow})\nonumber.
\end{align}
where $\varepsilon^{\mathrm{unif}}_{\mathrm{x}}(\mathrm{n})$ is the exchange energy per electron for the uniform electron gas and $F_{\mathrm{xc}}$ is an enhancement factor. 

By including $\tau$, the meta-GGA functional becomes more flexible and allows for satisfying a greater number of exact constraints compared to GGAs. Furthermore, by defining a dimensionless variable $\mathrm{\alpha} = \frac{ {\mathrm{\tau}-\mathrm{\tau}^{\mathrm{w}}} }{ {\mathrm{\tau}^{\mathrm{unif}}}  }$, where $\mathrm{\tau}^{\mathrm{unif}} = (3/10)(3\pi^2)^{2/3}\mathrm{n}^{5/3}$ is the kinetic energy density of the uniform electron gas and $\mathrm{\tau}^{\mathrm{w}} = |\nabla \mathrm{n}|^2/8\mathrm{n}$ is the von Weizs$\mathrm{\ddot{a}}$cker kinetic energy density, a meta-GGA can recognize slowly varying densities, single-orbital systems, and non-covalent bonds between two closed shells \cite{sun2015strongly,sun2016accurate,Sun2013}. Moreover, since $\tau$ is determined from the set of KS orbitals, meta-GGAs with explicit $\tau$ dependence are intrinsically non-local in nature. DFAs in this class include SCAN,  Regularized SCAN (rSCAN)\cite{bartok2019regularized}, regularized-restored SCAN ($\mathrm{r^{2}}$SCAN) \cite{furness2020accurate,furness2021construction}, Tao-Perdew-Staroverov-Scuseria (TPSS) \cite{tao2003climbing}, revised-TPSS (revTPSS)\cite{perdew2009workhorse}, meta-GGA made simple 0 (MS0) \cite{sun2012communication}, meta-GGA made simple 2 (MS2) \cite{sun2013semilocal}, and the Minnesota functional (M06L) \cite{zhao2006new}. 
 
 Trickey {\it et al.} recently substituted functions of $\nabla^2\mathrm{n}{(\mathbf{r})}$  for $\tau(\mathbf{r})$ in meta-GGA XC functionals, leading to SCAN-L \cite{mejia2017deorbitalization,mejia2018deorbitalized} and $\mathrm{r^{2}}$SCAN-L\cite{mejia2020meta} XC density functionals, among others, which yield similar (but not identical) performance to the original orbital dependent versions. The last rung in Perdew-Schmidt considered in this article are the hybrid functionals, which were originally designed to combine a semi-local DFA with the exact exchange of the Hartree--Fock single determinant to improve predictions of molecular thermochemical properties. The idea is that since the semi-local DFAs typically overbind while Hartree--Fock underbinds, their combination would presumably capture the correct balance between the two limits. The XC energy for the screened hybrid functional of Heyd, Scuseria, and Ernzerhof (HSE)  is given by 
\begin{align}
\mathrm{E_{xc}^{HSE}} = a \mathrm{E_{x}^{HF,SR}}(\omega) +(1-a)&\mathrm{E_{x}^{PBE,SR}}(\omega) \\  &+\mathrm{E_{x}^{PBE,LR}(\omega)+E_{c}^{PBE}},\nonumber
\end{align}
where \(a\) is the exact exchange admixing parameter whose typical value is $1/4$ \cite{krukau2006influence}. Here, the screening parameter $\omega$ defines the separation range, $\mathrm{E_{x}^{HF,SR}(\omega)}$ the short-range HF exact exchange, and $\mathrm{E_{x}^{PBE,SR}(\omega)}$ and $\mathrm{E_{x}^{PBE,LR}(\omega)}$ the short and long-range components, respectively, of the PBE exchange functional. The admixing parameter value of $a = 1/4$ has been justified through a consideration of molecular thermochemical properties \cite{perdew1996rationale}.

\begin{figure}[h] 
\includegraphics[width=1\linewidth]{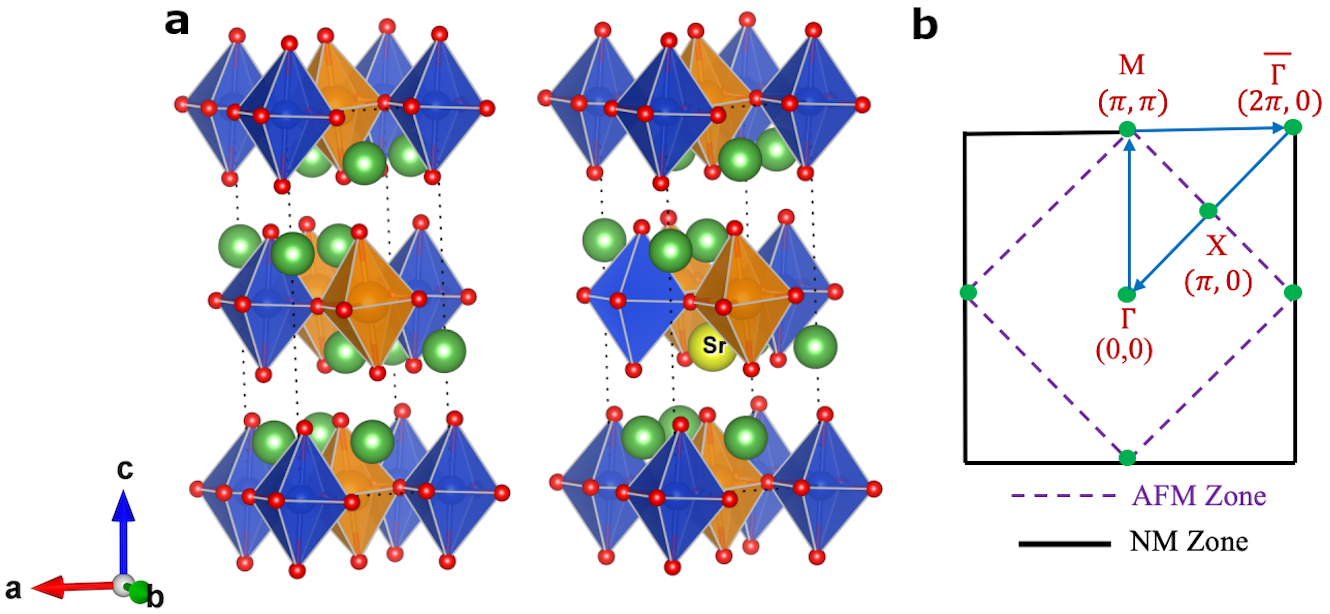} 
\caption{(\textbf{a}) Theoretically predicted crystal structure of  $\mathrm{La}_{2-x}\mathrm{Sr}_{x}\mathrm{CuO}_{4}$  in the LTO phase for $x=0.0$ and $0.25$. Copper, oxygen, lanthanum, and strontium atoms are represented by blue, red, green, and yellow spheres, respectively. Octahedral faces are shaded in blue (orange) to denote spin-up (down). Black dotted lines mark the unit cell. (\textbf{b})  A schematic of the non-magnetic (NM) and anti-ferromagnetic (AFM) Brillouin zone, where the path followed in the electronic dispersions in FIG.\ref{fig:band_lto} is marked.}
\label{fig:lto_t}
\end{figure}

\subsection*{Ground State Crystal Structure}
The phase diagram of the cuprates displays a complex intertwining of magnetic and charge ordered states that evolve with doping to reveal a superconducting dome. Interestingly, structural phase transitions associated with various octahedral tilt modes\cite{vaknin1987antiferromagnetism, freltoft1988magnetic} mainly follow the electronic phase boundaries.\cite{askerzade2012physical} At high temperatures LCO is found to be tetragonal (HTT) with all $\mathrm{CuO_{6}}$ octahedra aligned axially. A phase transition occurs upon lowering the temperature resulting in a low-temperature orthorhombic (LTO) phase where the octahedra are tilted along the (110) zone diagonal. An additional low-temperature tetragonal (LTT) phase arises upon substituting La with Ba or Nd, where the octahedral tilts are aligned along the (100) and (010) directions in alternating CuO$_2$ layers. Therefore, in order to properly disentangle the connection between the electronic and the physical properties of the cuprates, it is imperative to capture the correct ground state crystal structure. To calculate the total energies of various crystalline phases, we consider the $\sqrt{2}$ $\times$ $\sqrt{2}$ supercell of the body-centered-tetragonal I4/mmm primitive unit-cell to accommodate both the octahedral tilts and the $(\pi,\pi)$ AFM order within the CuO$_2$ planes. We treat the doping within a relatively simple \enquote{$\delta$- doping} scheme in which one La atom in the supercell is replaced by a Sr atom to yield an average hole doping of 25\%\cite{furness2018accurate}. This approach has been recently used for doping LSCO via molecular beam epitaxy techniques\cite{suter2018superconductivity}.  Figure \ref{fig:lto_t} (\textbf{a}) shows the crystal structures of LCO and LSCO in the LTO phase where the CuO$_6$ octahedra have been shaded blue and orange to represent the AFM order. The Sr doping site is also indicated.

\begin{figure*}[t] 
\includegraphics[width=0.99\linewidth]{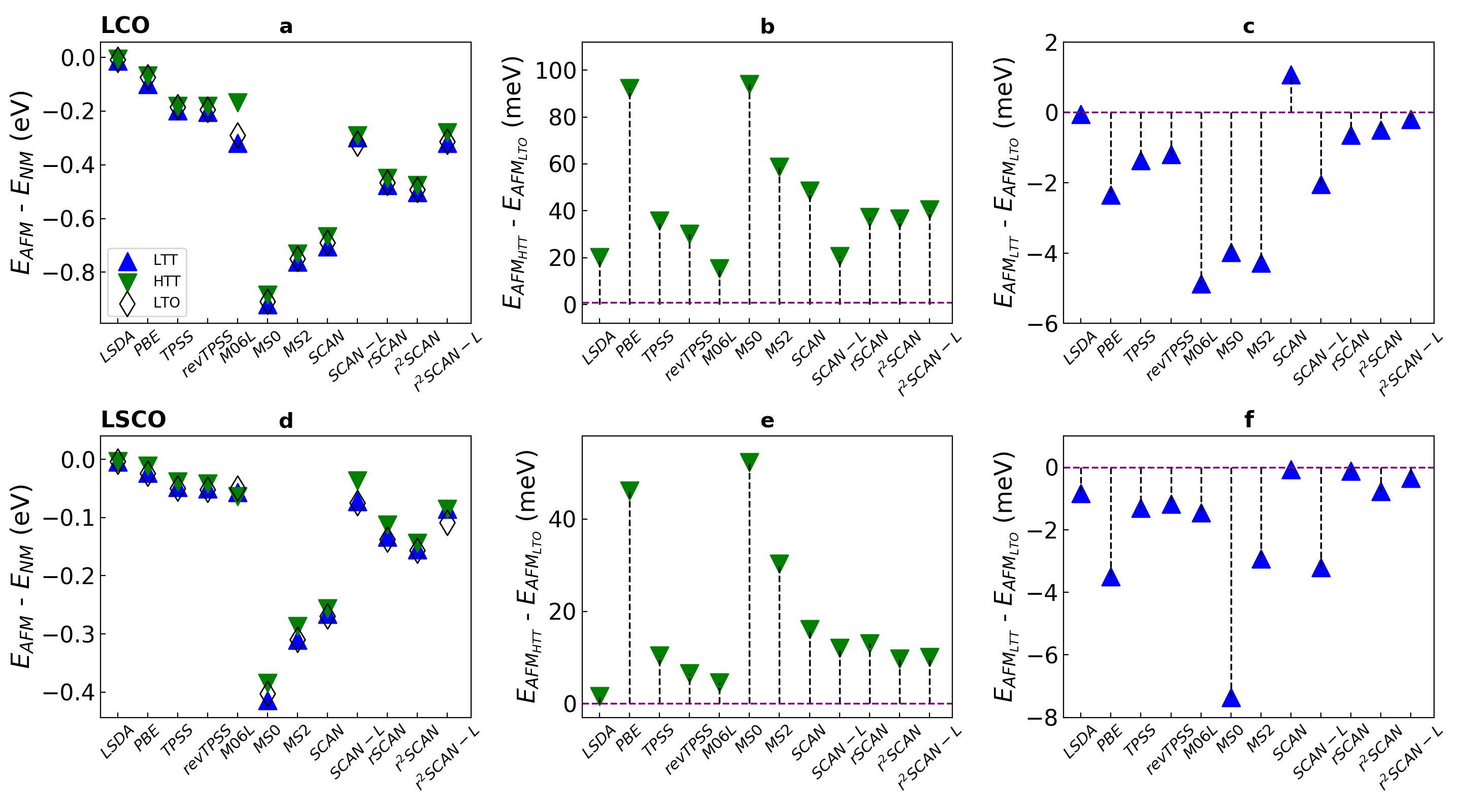} 
\caption{$(\textbf{a})$ Energy differences between the G-AFM and NM phases for the HTT (green upside-down triangle), LTO (white diamond), and LTT (blue triangle) structures for various XC density functionals. $(\textbf{b}$-$\textbf{c})$ Relative energies per formula unit for AFM in pristine LCO between LTO and HTT $(\textbf{b})$ and LTO and LTT $(\textbf{c})$. $(\textbf{d}$- $\textbf{f})$ Same as $(\textbf{a}$-$\textbf{c})$ except that these panels refer to  LSCO instead of LCO.}
\label{fig:energy}
\end{figure*}

\begin{figure}[h!] 
\includegraphics[width=\linewidth]{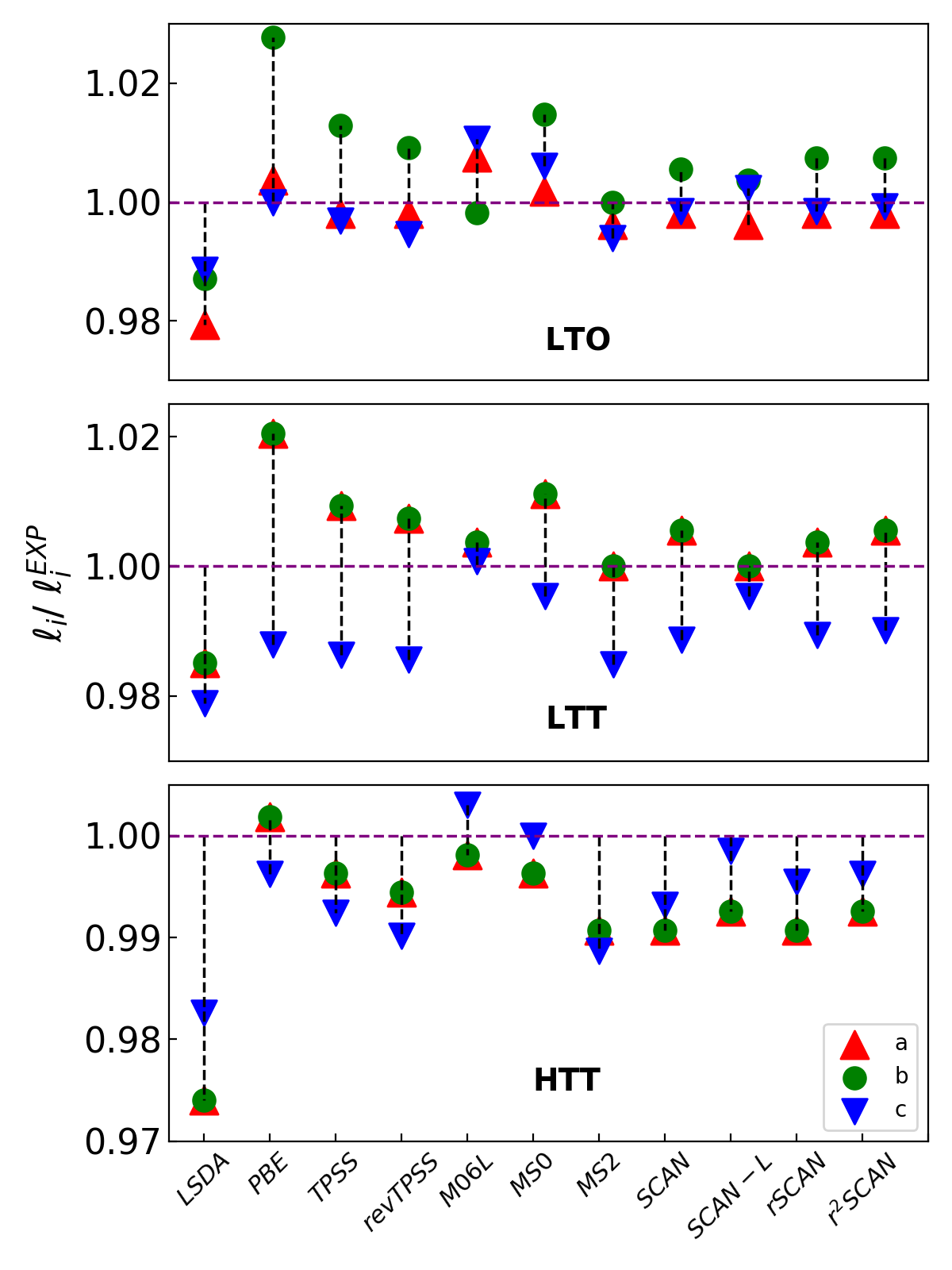} 
\caption{Comparison of the theoretically obtained and experimental lattice constants a, b, c for the HTT, LTT, and LTO crystal structures using various density functional approximations for La$_2$CuO$_4$. The lattice constant values are divided by corresponding experimental values.}
\label{fig:pristine_l}
\end{figure}

Figures \ref{fig:energy} (a) and \ref{fig:energy}(d) present energy differences between the AFM and NM phases for the pristine and doped $\mathrm{La}_{2-x}\mathrm{Sr}_{x}\mathrm{CuO}_{4}$ in each crystal structure for the DFAs considered. Firstly, we note that LSDA does not stabilize an AFM order over the Cu sites, whereas in PBE the AFM phase is marginally more stable, consistent with previous studies\cite{furness2018accurate}. All meta-GGAs find the AFM phase to be the ground state, with an energy separation of $-0.2$ to $-0.9$ eV between the AFM and NM states in the pristine structure, whereas in the doped case the energy difference is smaller by a factor of two. These trends are consistent across the various crystal structures.

Figures \ref{fig:energy} (b-c) and \ref{fig:energy}(e-f) present energy differences between the HTT, LTT, and LTO crystal structures for pristine and doped $\mathrm{La}_{2-x}\mathrm{Sr}_{x}\mathrm{CuO}_{4}$ for various density functional approximations. In all cases, the HTT phase lies at much higher energy compared to the LTO and LTT phases. Difference between the LTO and LTT appears more delicate. For the undoped case, only SCAN correctly predicts LTO to be the ground state, while LSDA, rSCAN, $\mathrm{r^{2}}$SCAN, and $\mathrm{r^{2}}$SCAN-L find LTO and LTT to be nearly degenerate with an energy difference of less than 1 meV. In the doped case, all XC functionals correctly predict the ground state to be LTT \cite{tranquada1995evidence}, while SCAN and rSCAN yield a marginal energy difference between LTT and LTO. Note that near 12\% doping, the LTO and LTT phases are found experimentally to be virtually degenerate \cite{furness2018accurate}.

Figure \ref{fig:pristine_l} shows the equilibrium lattice constants for LCO in the HTT, LTT, and LTO phases. The LSDA and PBE values were taken from Ref. \cite{furness2018accurate} and experimental values  from Refs. \cite{jorgensen1988superconducting, cox1989structural,wolf2012novel}. The LSDA is seen to underestimate the lattice constant for all crystal structures. PBE, on the other hand, underbinds the atoms and yields an exaggerated orthorhombicity in the LTO phase, similar to the super-tetragonality spuriously predicted by PBE for ferroelectric materials \cite{zhang2017comparative}. TPSS, revTPSS, MS0, MS2, SCAN, SCAN-L, rSCAN, $\mathrm{r^{2}}$SCAN and $\mathrm{r^{2}}$SCAN-L correct PBE by reducing the $b$ lattice constant in line with the experimental values in LTO and LTT. 

\begin{figure}[t] 
\includegraphics[width=\linewidth]{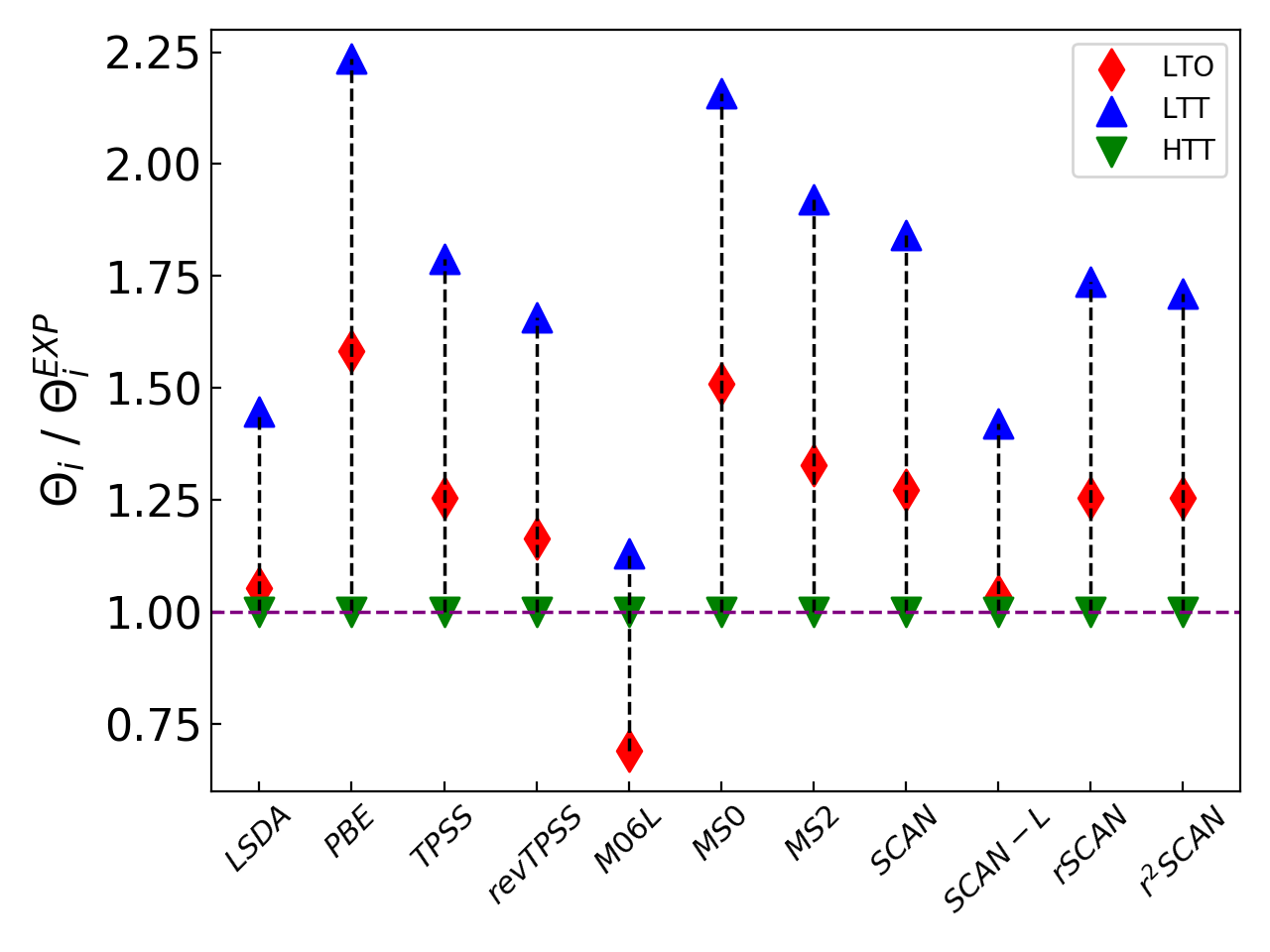} 
\caption{Theoretically predicted values of octahedra tilt angle using various density functional approximations for LCO. The LSDA and PBE value are taken from reference \cite{furness2018accurate} The octahedra tilt values for LTO, LTT and HTT are divided by corresponding experimental values.}
\label{fig:tilt}
\end{figure}

Curiously, all XC density functionals underestimate the lattice parameters in the HTT phase, except for PBE and M06L. The empirical M06L XC functional predicts lattice constants with greater accuracy than other XC functionals in all cases. Note that HTT is a high-temperature phase and therefore the experimental lattice constant should, in principle,  be corrected for finite-temperature effects for comparison with DFT results. Figure \ref{fig:tilt} considers the octahedral tilt angles. Here, M06L underestimates the tilt angle, while  all other XC functionals  overestimate it within a few degrees. We note, however, that the experimental tilt angles should be regarded as average values because the CuO$_6$ octahedra are not rigid objects: these octahedra couple to various phonon modes and deform dynamically. Molecular dynamics or phonon calculations will be needed to capture the octahedral tilts more accurately.

Lattice constants and octahedral tilts are not included for $\mathrm{r}^{2}$SCAN-L in Figs. \ref{fig:pristine_l} and \ref{fig:tilt} because we found a non-zero stress tensor at the energy-minimized equilibrium volume in this case. This suggests an error in the stress tensor implementation of $\mathrm{r}^{2}$SCAN-L, See Section S3 of Supplementary Materials for more details. The experimental structures were therefore used for the electronic and magnetic properties calculations using $\mathrm{r}^{2}$SCAN-L . The experimental structures were also used for HSE06 based calculations as the computational cost for hybrid XC functionals is much greater than the meta-GGAs.

Notably, within the “SCAN family” of XC density functionals (SCAN, SCAN-L, rSCAN, and $\mathrm{r}^{2}$SCAN) all members show similar performance for lattice constants and tilt angles (Figs.\ref{fig:pristine_l} and \ref{fig:tilt}). The potential speedup of running $\mathrm{r}^{2}$SCAN-L in a density-only KS scheme and the improved numerical performance inherited from $\mathrm{r}^{2}$SCAN suggest that $\mathrm{r}^{2}$SCAN-L could be used to optimize geometry followed by a single point SCAN or $\mathrm{r}^{2}$SCAN calculation for obtaining electronic properties. This approach may also present an advantage of minimizing the numerical challenges associated with SCAN. A similar scheme was suggested in Ref. \cite{mejia2020spin} in the context of spin-crossover prediction.

\begin{figure}[t!] 
\centering
\includegraphics[width=\linewidth]{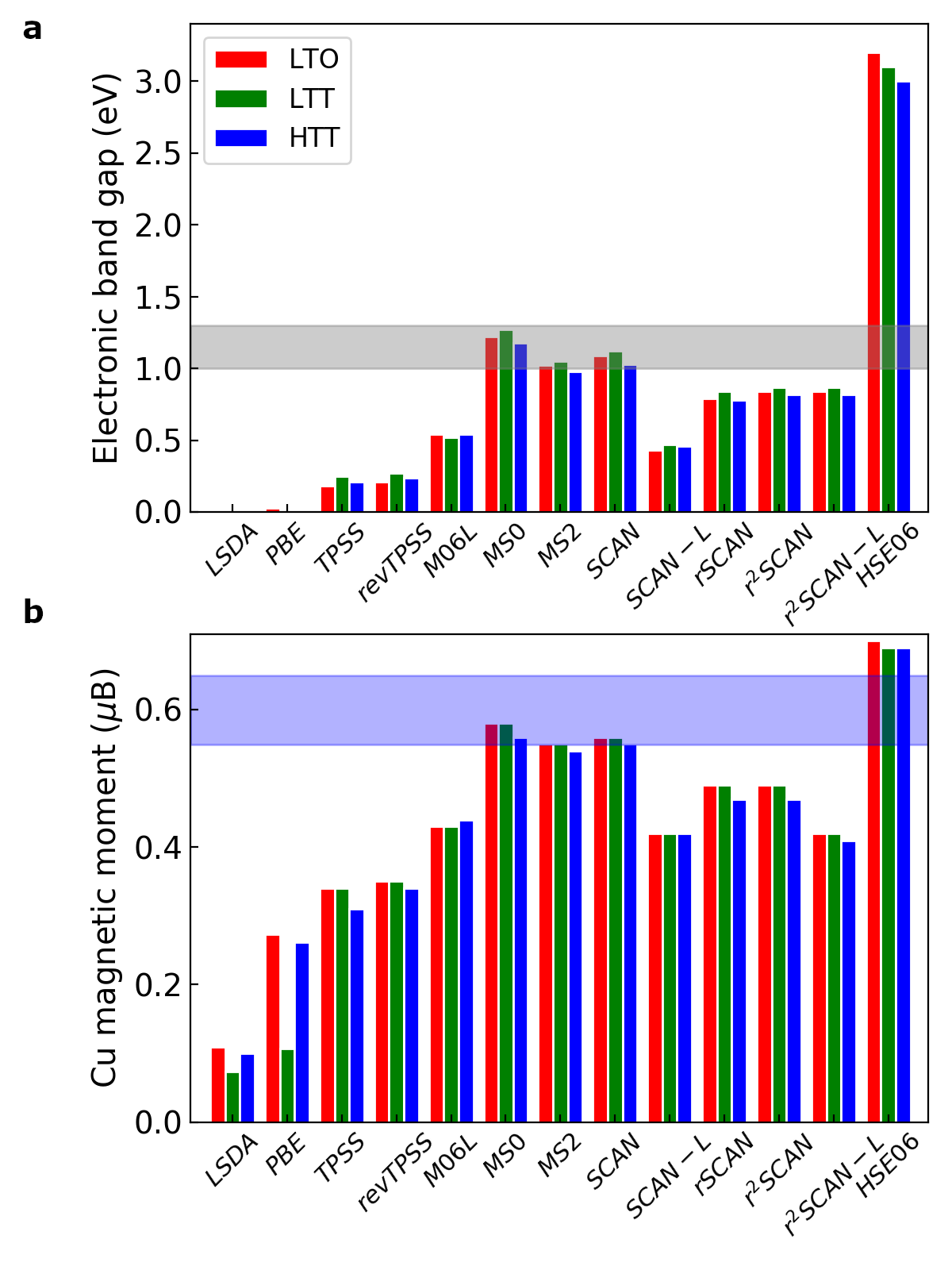} 
\caption{Theoretical predicted values of (\textbf{a})  electronic band gap and (\textbf{b}) copper magnetic moment for all three phases of pristine LCO obtained within various density functional approximations.  The gray shaded region in (\textbf{a}) gives the spread in the reported experimental values for the leading edge gap \cite{uchida1991optical,li2012optical,kastner1998magnetic}. In (\textbf{b}), the blue shaded region represents the experimental value of magnetic moment \cite{tranquada2007handbook}. The LSDA and PBE values are taken from Ref. \cite{furness2018accurate}} 
\label{fig:band_mag}
\end{figure}

\subsection*{Electronic and Magnetic Structures}

Figure~\ref{fig:band_mag}  compares the theoretically predicted electronic bandgaps and copper magnetic moments obtained from various XC functionals for the three crystalline phases of LCO.
The range of experimentally observed bandgaps \cite{uchida1991optical,li2012optical, kastner1998magnetic}, and median copper magnetic moments\cite{tranquada2007handbook} are marked by the grey and blue shaded regions, respectively. LSDA and PBE greatly underestimate the bandgaps and magnetic moments because they fail to stabilize the AFM order. A large variation is seen in the results of the meta-GGAs. TPSS and revTPSS both  underestimate the bandgaps and magnetic moments. MS0, MS2 and SCAN yield values that lie within the experimental ranges. Other meta-GGAs predict reduced bandgaps and magnetic moments that are below experimental values. M06L underestimates both the moment and the bandgap value, possibly due to the bias towards molecular systems which is encoded in its empirical construction. M06L yields ferrimagnetic order and, therefore, the average of the magnetic moment is given in Fig. \ref{fig:band_mag}. Finally, the hybrid functional (HSE06) overestimates bandgaps, predicting a value of around $3$ eV, and it also overestimates magnetic moments.

Ando\cite{ono2007strong} has stressed that one should estimate the bandgap  not from the lowest energy absorption peak, but from the leading edge gap in the optical spectra\cite{uchida1991optical}. The leading edge gives the minimum energy needed by an electron to be elevated from the valence to the conduction band, in good agreement with the transport gap in the cuprates.  In contrast, the energy of the absorption peak in the optical spectrum depends on finer details of the electronic structure such as the presence of flat bands or Van Hove singularities. The theoretically predicted bandgaps here should be compared to the fundamental band gaps \cite{cohen2008fractional, mori2008localization, perdew2017understanding, zhang2020symmetry}, which are typically larger than the corresponding optical band gaps due to excitonic effects. Notably, a recent measurement on LCO reports an optical bandgap of about $\approx$ 1.3 eV \cite{li2012optical}.

Regarding magnetic moments, the values obtained by neutron scattering involve uncertainties since the copper form factor is not {\it a priori} known. Appendix E of Ref.[\onlinecite{lane2018antiferromagnetic}] compares copper magnetic moments from various experiments, including the values given in the recent review of Tranquada \cite{tranquada2007handbook}. Note that, when estimating the copper magnetic moment, we have increased the Wigner-Seitz radius of the integration sphere beyond the default 1.16 \AA~  to 1.91 \AA~ (the Cu-O bond length) in order to fully capture the magnetic density centered on the copper atomic site and the part originating from strong hybridization between the copper and oxygen atoms (see Section S4 in the Supplementary materials for more details).

While these predicted magnetic moments are in approximate agreement with the experimentally measured value, it has been suggested that this is not the correct comparison. This is because 1) given a static magnetic moment, fluctuation can cause a ~30\% reduction \cite{auerbach2012interacting}. 2) DFT works with spin symmetry breaking \cite{perdew2021interpretations}, and the DFT magnetic moments should be compared to the static value. We believe this issue is unresolved and that self-consistency can capture some average effects of fluctuation due to the DFT correlation potential. We hope to discuss this further in a future publication.

\begin{figure}[t!] 
\includegraphics[width=\linewidth]{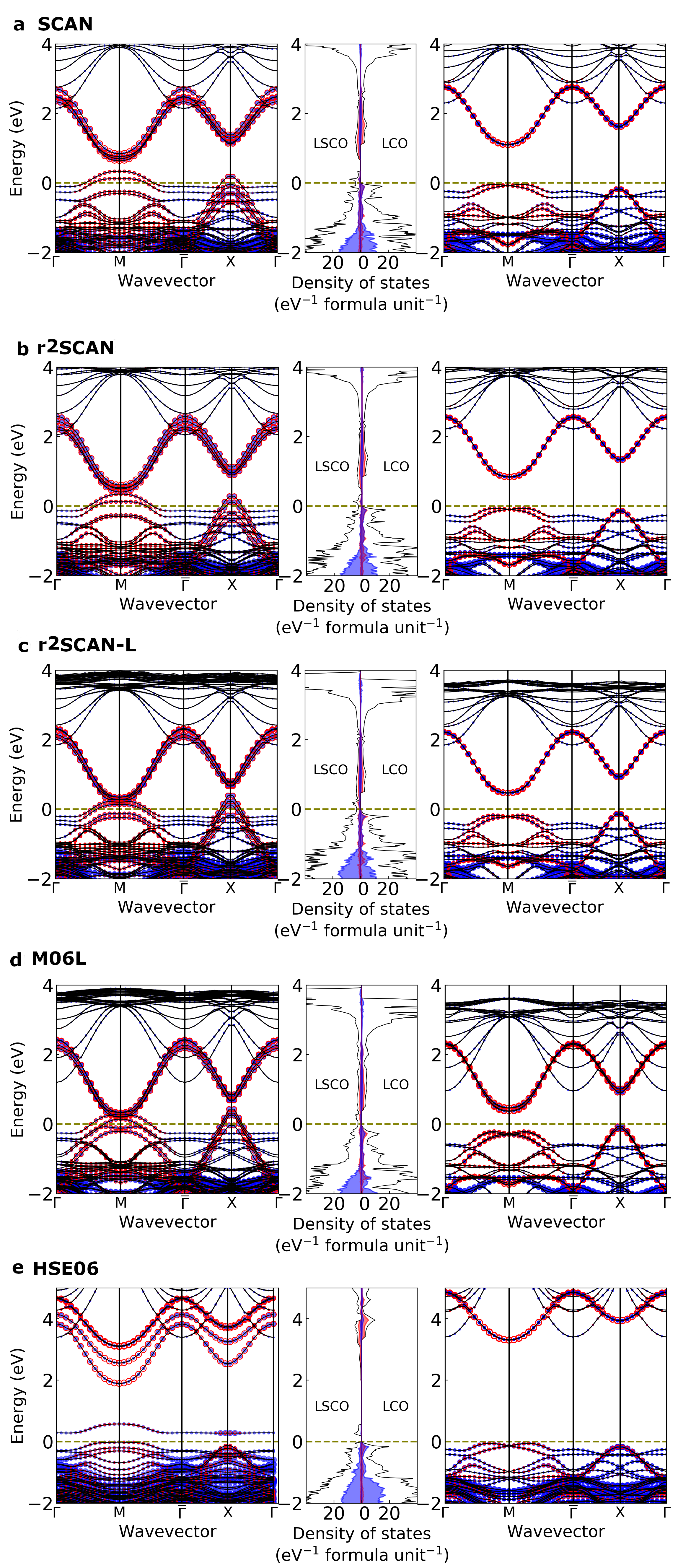} 
\caption{Electronic band structure and density of states of LCO and LSCO in the LTO phase using $(\textbf{a})$ SCAN  $(\textbf{b})$ $\mathrm{r^{2}}$SCAN $(\textbf{c})$ $\mathrm{r^{2}}$SCAN-L $(\textbf{d})$ M06L $(\textbf{e})$ HSE06. The contribution of Cu-$\mathrm{d}_{\mathrm{x^2}-\mathrm{y^2}}$ and O -$\mathrm{p_x}+\mathrm{p_y}$ are marked by the red and blues shadings, respectively. The path followed by the dispersion in the Brillouin zone is shown in Fig \ref{fig:lto_t}$(\textbf{b})$. }
\label{fig:band_lto}
\end{figure}

Figure \ref{fig:band_lto} presents the electronic band dispersions in pristine and doped La$_2$CuO$_4$ in the LTO crystal structure for the AFM phase using SCAN, $\mathrm{r^{2}}$SCAN, $\mathrm{r^{2}}$SCAN-L, M06L, and HSE06. The copper (red circles) and planar oxygen (blue dots) orbital contributions are overlayed. For all XC functionals, LCO is seen to be an insulator. At the valence band edge, SCAN, $\mathrm{r^{2}}$SCAN and $\mathrm{r^{2}}$SCAN-L produce a significant avoided crossing between the $\mathrm{d}_{\mathrm{x^2}-\mathrm{y^2}}$ and in-plane oxygen dominated bands along $\Gamma-\mathrm{M}$ and $\mathrm{M}-\bar{\Gamma}$, but this feature is essentially absent in M06L. In SCAN and $\mathrm{r^{2}}$SCAN, the gap is direct, with its smallest value occurring at $\mathrm{M}$ symmetry point or very close to it. In contrast, $\mathrm{r^{2}}$SCAN-L and M06L predict indirect bandgaps. Finally, the conduction bands in M06L display significant spin splittings indicative of  ferrimagnetic ordering consistent with the observed ferrimagnetic  moments.

Turning to the doped system in Fig.\ref{fig:band_lto}, all meta-GGAs are seen to capture the metal-insulator transition, with various XC functionals producing small differences in band splittings around the Fermi level. In contrast, HSE06 maintains a small gap and predicts a nearly flat impurity-like band just above the Fermi level, consistent with the B3LYP results \cite{perry2002ab}. See Supplementary Materials for further details of the electronic band dispersions in LTO, LTT and HTT phases. The SCAN-based magnetic moments and bandgaps given in this study differ by $\sim 0.02$ ${\mathrm{\mu}_B}$ and 0.11eV respectively, from those given in Ref.\cite{furness2018accurate}. These small differences, which do not affect the overall conclusions of Ref.\cite{furness2018accurate}, are due to an error in the VASP implementation that was used in Ref. \cite{furness2018accurate}.

\subsection*{Effective U and exchange coupling}
The band gap that develops in the half-filled Cu $\mathrm{d}_{x^2-y^2}$ dominated band by splitting the up- and down-spin bands is due to strong multi-orbital intrasite electron-electron interactions. The strength of these interactions is a key quantity that can be used to  characterize various regions of the phase diagram and classify the phenomenology of the cuprate family as a whole\cite{markiewicz2017entropic}. In order to estimate the correlation strengths implicit in the underlying XC density functionals, we map our site-resolved partial densities of states to a multiorbital Hubbard model\cite{oles1983antiferromagnetism} along the lines of  Ref. \cite{lane2018antiferromagnetic}. For this purpose, we consider a $\mathrm{d}$ orbital $\mathrm{\mu}$ of spin $\mathrm{\sigma}$ in a ligand field with on-site correlations in the mean field, and express its energy as 
\begin{align}\label{eq:0}
    \mathrm{E}^{\mathrm{\mu\sigma}}_{\pm} &=
     \mathrm{E}^{\mathrm{\mu}}_{\text{atomic}} 
    +  \mathrm{U} \braket{n_{\mathrm{\mu\bar{\sigma}}}^{\pm} }
    + \mathrm{U}^{\prime}\sum_{\nu\neq \mu}\braket{n_{\nu\bar{\sigma}}^{\pm} }\\
    &+ (\mathrm{U}^{\prime}-\mathrm{J_{H}}) \sum_{\nu\neq \mu}\braket{n_{\nu{\sigma}}^{\pm}}
    \pm \mathrm{h},\nonumber
\end{align}
where $\pm$ indexes the bonding $(-)$ and antibonding $(+)$
states, and $\mathrm{h}$ is the hybridization strength. $\mu(\nu)$ and spin $\sigma (\bar{\sigma}=-\sigma)$ are orbital and spin indices, respectively, and $\braket{n^{\pm}_{\mu \sigma}}$ is the average electron occupation for a given state in the mean field. By taking the difference between the up- and down-spin channels and summing over bonding and anti-boding levels, $\mathrm{U}$ and $\mathrm{J_{H}}$ can be shown to  connect the spin splitting of a given orbital to the differences in various spin-dependent orbital occupations,
\begin{equation}\label{eq:1}
\mathrm{E}^{{\mathrm{\mu}}{\uparrow}} - \mathrm{E}^{{\mathrm{\mu}}{\downarrow}} = \mathrm{U}(\mathrm{N}_{{\mathrm{\mu}}{\downarrow}} - \mathrm{N}_{{\mathrm{\mu}}{\uparrow}}) - \\\mathrm{J}_{\mathrm{H}}\sum_{{\mathrm{\nu}}{\neq}{\mathrm{\mu}}}(\mathrm{N}_{{\mathrm{\nu}}{\uparrow}} - \mathrm{N}_{{\mathrm{\nu}}{\downarrow}}),
\end{equation}
where  $\mathrm{N}_{{\mathrm{\mu}}{\mathrm{\sigma}}} =\sum_{\pm}{\braket{n^{\pm}_{\mu \sigma}}}$.  Furthermore, $\mathrm{E}^{{\mathrm{\mu}}{\mathrm{\sigma}}}$ may be obtained from the  density of states:
\begin{equation}\label{eq:2}
\mathrm{E}^{{\mathrm{\mu}}{\mathrm{\sigma}}} = \int_{W} \mathrm{g}_{{\mathrm{\mu}}{\mathrm{\sigma}}}\mathrm{(\varepsilon)\varepsilon d\varepsilon}   
\end{equation}
where $W$ represents the bandwidth. The average spin splitting of a given orbital can then be expressed as:
\begin{equation}\label{eq:3}
\mathrm{E}^{{\mathrm{\mu}}{\uparrow}}-\mathrm{E}^{{\mathrm{\mu}}{\downarrow}} =\int_{W} [\mathrm{g}_{{\mathrm{\mu}}{\uparrow}}\mathrm{(\varepsilon)}- \mathrm{g}_{{\mathrm{\mu}}{\downarrow}}\mathrm{(\varepsilon)]\varepsilon d\varepsilon}. \end{equation}
We thus arrive at the following coupled set of equations for the interaction parameters,
\begin{align}\label{eq:4}
\int_{W}& [\mathrm{g}_{{\mathrm{\mu}}{\uparrow}}\mathrm{(\varepsilon)}- \mathrm{g}_{{\mathrm{\mu}}{\downarrow}}\mathrm{(\varepsilon)]\varepsilon d\varepsilon} =\\ &\mathrm{U}(\mathrm{N}_{{\mu\downarrow}} - \mathrm{N}_{{\mu\uparrow}}) - \mathrm{J}_{\mathrm{H}}\sum_{{\mathrm{\nu}}{\neq}{\mathrm{\mu}}}(\mathrm{N}_{{\mathrm{\nu}}{\uparrow}} - \mathrm{N}_{{\mathrm{\nu}}{\downarrow}}).\nonumber
\end{align}

By using the copper-atom-projected partial density-of-states in the AFM phase of LTO La$_2$CuO$_4$ where the $\mathrm{d}_{\mathrm{x^2-y^2}}$ orbital is half-filled and all other orbitals are completely filled,  we can simplify the preceding set of coupled equations into the form:
\begin{equation}
\int_{W} [\mathrm{g}_{{\mathrm{d}}_{\mathrm{x^2}-\mathrm{y^2}}{\uparrow}}\mathrm{(\varepsilon)}- \mathrm{g}_{{\mathrm{d}}_{\mathrm{x^2}-\mathrm{y^2}}{\downarrow}}\mathrm{(\varepsilon)]\varepsilon d\varepsilon}= \mathrm{U/2}
\end{equation}
\begin{equation}
\int_{W} [\mathrm{g}_{({\mathrm{\mu}}{\neq}\mathrm{d}_{\mathrm{x^2}-\mathrm{y^2}}){\uparrow}}\mathrm{(\varepsilon)}-\mathrm{g}_{({\mathrm{\mu}}{\neq}\mathrm{d}_{\mathrm{x^2}-\mathrm{y^2}}){\downarrow}}\mathrm{(\varepsilon)]\varepsilon d\varepsilon} = \mathrm{J_{H}/2}
\end{equation}

Finally, we evaluate $\int_{W}[\mathrm{g}_{{\mathrm{\mu}}{\uparrow}}\mathrm{(\varepsilon)}-\mathrm{g}_{{\mathrm{\mu}}{\downarrow}}\mathrm{(\varepsilon)]\varepsilon d\varepsilon}$ over the full band width $W$ for each orbital to solve for $\mathrm{U}$ and $\mathrm{J_H}$. The estimated values of $\mathrm{U}$ and $\mathrm{J_H}$ so obtained are presented in Table~\ref{table:u}. The average spin-splittings are strongly orbital dependent \cite{lane2018antiferromagnetic}, and we have taken the largest value as the upper bound on $\mathrm{J_H}$.

\begin{table}[h!]
\caption{Theoretically  predicted values of U and $\mathrm{J_{H}}$ using various DFAs for three different phases of pristine LCO. }
\begin{tabular}{lc|c|cc}
\hline
 & Functional & Phase & U (eV)&$\mathrm{J_{H}}$ (eV) \\ 
\hline\hline
& & LTO&2.23&0.54\\
& {TPSS} & LTT&2.19&0.55\\
& & HTT &2.19&0.54\\
\hline\hline
& & LTO&2.32&0.60\\
& {revTPSS} & LTT&2.31&0.60\\
& & HTT &2.3&0.58\\
\hline\hline
& & LTO&3.14&0.51\\
& {M06L} & LTT&3.14&0.54\\
& & HTT &3.19&0.55\\
\hline\hline
& & LTO&5.60&1.36\\
& {MS0} & LTT&5.71&1.32\\
& & HTT &5.91&1.34\\
\hline\hline
& & LTO&5.00&1.16\\
& {MS2} & LTT&5.09&1.13\\
& & HTT &5.108&1.18\\
\hline\hline
& & LTO&5.40&1.25\\
& {SCAN} & LTT&5.40&1.27\\
& & HTT &5.36&1.24\\
\hline\hline
& & LTO&3.13&0.61\\
& {SCAN-L} & LTT&3.13&0.61\\
& & HTT &3.16&0.60\\
\hline\hline
& & LTO&4.24&1.04\\
& {rSCAN} & LTT&4.25&1.03\\
& & HTT &4.26&1.02\\
\hline\hline
& & LTO&4.45&1.06\\
& {$\mathrm{r^{2}}$SCAN} & LTT&4.43&1.06\\
& & HTT &4.41&1.05\\
\hline\hline
& & LTO&3.14&0.61\\
& {$\mathrm{r^{2}}$SCAN-L} & LTT&3.15&0.62\\
& & HTT &3.16&0.61\\
\hline\hline
& & LTO&11.79&1.27\\
& {HSE06} & LTT&11.30&1.36\\
& & HTT &11.58&1.27\\
 \hline
 \end{tabular}
 
 \label{table:u}
\end{table}
Results of Table \ref{table:u} show that TPSS and revTPSS yield a smaller value for $\mathrm{U}$ compared to the recent cRPA calculations ($\sim$ 3.2 eV) \cite{jang2016}, since they fail to adequately capture the bandgaps and magnetic moments, while M06L, SCAN-L and $\mathrm{r^{2}}$SCAN-L yield comparable values. MS0,MS2, rSCAN and $\mathrm{r^{2}}$SCAN, find larger values than the cRPA values. The hybrid HSE06 XC functional  predicts exaggerated values for $\mathrm{U}$. 

In order to determine the exchange coupling strength, we use a mean-field approach, where we map the total energies of the AFM and ferromagnetic (FM) phases onto those of a nearest-neighbor spin$-\frac{1}{2}$ Heisenberg Hamiltonian \cite{su1999crystal,coffey1990effective,noodleman1981valence}. The difference in the energies of the AFM and FM phases in the mean field limit is given by
\begin{equation}
\Delta \mathrm{E} =  \mathrm{E}_{\mathrm{AFM}}-\mathrm{E}_{\mathrm{FM}} = \mathrm{JNZ} {{\textless}\mathrm{S}{\textgreater}}^2   
\end{equation}
where N is the total number of magnetic sites in the unit cell, S =$1/2$ is the spin on each site, and Z is the coordination number. Since the in-plane interactions within the Cu-O planes in $\mathrm{La_{2}CuO_{4}}$ are much stronger than the interplanar interactions, we take Z = 4. For our AFM $\sqrt{2} \times \sqrt{2}$ unit cell,
N = 4. In this way, we obtain the J values for various XC functionals listed in table \ref{table:coup}. 

\begin{table}[h!]
\caption{Theoretically  predicted values of exchange coupling using various XC functionals for three different phases of pristine LCO. }
\begin{tabular}{lc|c|c}
\hline
 & Functional & Phase & J (meV)  \\ 
\hline\hline
& & LTO&-26.74\\
& {TPSS} & LTT&-25.9\\
& & HTT &-22.24\\
\hline\hline
& & LTO&-26.89\\
& {revTPSS} & LTT&-27.47\\
& & HTT &-24.74\\
\hline\hline
& {M06L} &did not converge&- \\
\hline\hline
& & LTO&-158.29\\
& {MS0} & LTT&-159.36\\
& & HTT &-160.75\\
\hline\hline
& & LTO&-140.46\\
& {MS2} & LTT&-141.76\\
& & HTT &-139.94\\
\hline\hline
& & LTO&-131.08\\
& {SCAN} & LTT&-131.32\\
& & HTT &-125.97\\
\hline\hline
& & LTO&-48.48\\
& {SCAN-L} & LTT&-50.62\\
& & HTT &-49.95\\
\hline\hline
& & LTO&-87.16\\
& {rSCAN} & LTT&-88.37\\
& & HTT &-82.09\\
\hline\hline
& & LTO&-93.12\\
& {$\mathrm{r^{2}}$SCAN} & LTT&-95.04\\
& & HTT &-88.33\\
\hline\hline
& & LTO&-49.01\\
& {$\mathrm{r^{2}}$SCAN-L} & LTT&-50.61\\
& & HTT &-49.88\\
\hline\hline
& & LTO&-182.11\\
& {HSE06} & LTT&-188.51\\
& & HTT &-180.27\\

 \hline
 \end{tabular}
 \label{table:coup}
\end{table}

\begin{table}[h!]
\caption{Theoretically  predicted values  of the charge-transfer energies between Cu 3d and O 2p orbitals and two Cu energy splitting using various XC functionals  for three different phases of pristine LCO and doped LSCO systems }
\resizebox{\columnwidth}{!}{
\begin{tabular}{lc|c|cc|cc}
\hline
& & &\multicolumn{2}{c|}{ \textbf{pristine LCO}}& \multicolumn{2}{c}{\textbf{doped LSCO}} \\
 \hline
 & Functional & Phase & $\Delta_{\mathrm{dp}}$(eV) &$\Delta_{{\mathrm{e}}_{\mathrm{g}}}$(eV) &$\Delta_{\mathrm{dp}}$(eV) &$\Delta_{{\mathrm{e}}_{\mathrm{g}}}$(eV)  \\ 
\hline\hline
& & LTO&2.41&0.74&3.49&0.60\\
& {PBE} & LTT&2.38&0.75&3.30&0.62\\
& & HTT &2.41&0.79&3.46&0.59\\
\hline\hline
& & LTO&2.41&0.77&3.52&0.62\\
& {TPSS} & LTT&2.23&0.78&3.50&0.63\\
& & HTT &2.26&0.80&3.44&0.63\\
\hline\hline
& & LTO&2.37&0.77&3.50&0.61\\
& {revTPSS} & LTT&2.24&0.78&3.49&0.62\\
& & HTT &2.23&0.80&3.46&0.63\\
\hline\hline
& & LTO&2.52&1.00&3.95&0.73\\
& {M06L} & LTT&2.54&1.06&3.91&0.74\\
& & HTT &2.46&1.07&3.75&0.81\\
\hline\hline
& & LTO&2.99&1.34&5.00&1.07\\
& {MS0} & LTT&2.84&1.37&4.75&1.14\\
& & HTT &2.93&1.35&5.16&1.09\\
\hline\hline
& & LTO&3.00&1.21&4.76&0.89\\
& {MS2} & LTT&2.91&1.22&4.97&0.87\\
& & HTT &2.92&1.20&4.66&0.93\\
\hline\hline
& & LTO&3.01&1.23&4.84&0.95\\
& {SCAN} & LTT&2.93&1.24&4.79&0.95\\
& & HTT &2.92&1.24&4.71&0.96\\
\hline\hline
& & LTO&2.64&0.96&4.18&0.72\\
& {SCAN-L} & LTT&2.54&0.96&4.20&0.72\\
& & HTT &2.49&0.95&3.89&0.80\\
\hline\hline
& & LTO&2.55&1.06&4.17&0.92\\
& {rSCAN} & LTT&2.45&1.08&4.18&0.92\\
& & HTT &2.47&1.12&4.1&0.93\\
\hline\hline
& & LTO&2.52&1.08&4.19&0.92\\
& {$\mathrm{r^{2}}$SCAN} & LTT&2.45&1.09&4.22&0.93\\
& & HTT &2.47&1.13&4.15&0.93\\
\hline\hline
& & LTO&2.65&0.98&4.18&0.73\\
& {$\mathrm{r^{2}}$SCAN-L} & LTT&2.55&0.98&4.21&0.74\\
& & HTT &2.46&0.96&3.91&0.83\\
\hline\hline
& & LTO&7.35&2.76&9.89&2.80\\
& {HSE06} & LTT&6.91&2.82&-&-\\
& & HTT &7.15&2.74&-&-\\
\hline\hline
 \end{tabular}
 }
 \label{table:charge-transfer}
\end{table}
Table \ref{table:coup} shows that SCAN is most accurate in predicting the experimental value of $-133 \pm 3$ meV \cite{bourges1997superexchange,hayden1991high,coldea2001spin} for the exchange coupling in LCO. MS0 and MS2 slightly overestimate J compared to SCAN. TPSS, revTPSS, SCAN-L, rSCAN, $\mathrm{r^{2}}$SCAN and $\mathrm{r^{2}}$SCAN-L underestimate and HSE06 significantly overestimates J. M06L failed to converge for the FM case. Notably, here and in Ref. \cite{lane2018antiferromagnetic},our modeling involves only the nearest-neighbor J in keeping with the related experimental analysis.
We would expect some renormalization of the J values if we were to include next and higher nearest neighbors in our modeling.

In order to gain  further insight into the multiorbital nature of the electronic structure, two additional descriptors were estimated: (1) Charge-transfer energies between the Cu 3d and O 2p orbitals ($\Delta_{\mathrm{dp}}$); and (2) the tetragonal splitting of the $\mathrm{e_g}$ states  ($\Delta_{{\mathrm{e}}_{\mathrm{g}}}$), which are defined as
\begin{equation}
\Delta_{\mathrm{dp}} = \varepsilon_{\mathrm{d}}-\varepsilon_{\mathrm{p}}    
\end{equation}
and
\begin{equation}
\Delta_{{\mathrm{e}}_{\mathrm{g}}} = \varepsilon_{\mathrm{x^2}-\mathrm{y^2}}-\varepsilon_{\mathrm{z^2}}.   
\end{equation}
The various band centers $\mathrm{\varepsilon_{\mu}}$ are defined using the corresponding partial density-of-states as 
\begin{equation}
\mathrm{\varepsilon}_{\mathrm{\mu}} = \frac{\int \mathrm{g}_{\mathrm{\mu}}\mathrm{(\varepsilon)\varepsilon d\varepsilon}}{\int \mathrm{g}_{\mathrm{\mu}}\mathrm{(\varepsilon)d\varepsilon}}   , 
\label{eq:eq1}
\end{equation}
along the lines of Refs.\cite{botana2020similarities} and \cite{jang2015quasiparticle}. We used an integration window of $-8$ eV to the top of the band in Eq.\ref{eq:eq1}. This window covers only the anti-bonding bands for $\Delta_{{\mathrm{e}}_{\mathrm{g}}}$. Results of Table \ref{table:charge-transfer} show that compared to PBE, the meta-GGAs generally tend to enhance $\Delta_{\mathrm{dp}}$ and
$\Delta_{{\mathrm{e}}_{\mathrm{g}}}$ due to the stabilization of the AFM order. TPSS and revTPSS performances are comparable to PBE while other meta-GGAs predict larger  $\Delta_{\mathrm{dp}}$ and $\Delta_{{\mathrm{e}}_{\mathrm{g}}}$ values. For the doped case, $\Delta_{\mathrm{dp}}$ increases, whereas $\Delta_{{\mathrm{e}}_{\mathrm{g}}}$ reduces compared to the pristine results. HSE06 predicts significantly large $\Delta_{\mathrm{dp}}$ and $\Delta_{{\mathrm{e}}_{\mathrm{g}}}$  for both pristine and doped cases.

\subsection*{Meta-GGA performance}
The present results for the crystal, electronic, and magnetic properties clearly demonstrate that meta-GGAs  provide an improvement over LSDA and PBE. Among the various meta-GGAs considered (TPSS, revTPSS, MS0, MS2, SCAN, rSCAN, $\mathrm{r^{2}}$SCAN, M06L), M06L is less satisfactory for predicting the LCO properties, which is heavily parameterized for molecular systems. The earlier  generalized-KS (gKS) meta-GGAs such as TPSS and revTPSS are less accurate than the more recently developed approximations (e.g. SCAN). The success of SCAN is a consequence of its enforcing all the known 17 rigorous constraints that a semilocal functional can satisfy\cite{sun2015strongly}. In addition, SCAN localizes d electrons better by reducing self-interaction errors that tend to over-delocalize d electrons in the presence of ligands involving s and p electrons\cite{zhang2020symmetry}. SCAN thus stabilizes the magnetic moment of Cu and opens a sizable bandgap in LCO\cite{furness2018accurate}, its shortcomings in exaggerating magnetic moments in 3d elemental solids notwithstanding \cite{isaacs2018performance}.

rSCAN solves the numerical grid issues encountered in SCAN by regularizing the problematic interpolation function of SCAN with a smooth polynomial, which unfortunately violates exact constraints \cite{bartok2019regularized, furness2020accurate}, and some of rSCAN's tranferability is lost \cite{mejia2019comment, bartok2019response}. $\mathrm{r^{2}}$SCAN retains the smoothness of rSCAN and maximally restores the exact constraints violated by the regularization of rSCAN and it has been shown to improve the accuracy over rSCAN while maintaining the numerical efficiency\cite{furness2020accurate}. In the present study of cuprates, $\mathrm{r^{2}}$SCAN and rSCAN both perform similarly, with only slight underestimations of the band gaps and magnetic moments. 

By replacing the kinetic energy density with the Laplacian of the electron density and thus de-orbitalizing the underlying meta-GGAs, SCAN-L \cite{mejia2017deorbitalization} and $\mathrm{r^{2}}$SCAN-L \cite{mejia2020meta} are constructed from SCAN and $\mathrm{r^{2}}$SCAN, respectively. The XC potentials in SCAN-L and $\mathrm{r^{2}}$SCAN-L are locally multiplicative while in their orbital dependent parent functionals, the potentials are nonmultiplicative. Perdew {\it et al.}\cite{perdew2017understanding} have shown that for a given DFA, the gKS orbital band gap is equal to the corresponding fundamental band gap in solids, which is defined as the second order ground-state energy difference with respect to electron number. This indicates that within the gKS formalism a DFA with better total energy also improves the band gap \cite{zhang2020symmetry}. The preceding statement also applies to DFAs with multiplicative potentials as they have the same potentials in the KS and gKS schemes. The bandgaps and copper magnetic moments from SCAN-L and $\mathrm{r^{2}}$SCAN-L are consistently underestimated compared to the corresponding values from the parent SCAN and $\mathrm{r^{2}}$SCAN XC functionals.

M06L being an empirical functional is heavily parametrized. It is constructed by fitting to molecular data sets, and therefore, it tends to be less reliable for systems outside its fitting set with limited transferability.

\subsection*{Why does HSE06 open a gap in the doped LSCO?}
HSE06 with the admixing parameter value of 1/4 works well for bandgap predictions in semiconductors. This improvement is due to the reduction of the self-interaction error present in PBE through the introduction of exact exchange \cite{mori2008localization,zhang2020symmetry}. However, Hartree-Fock is not applicable to metallic systems where there is no bandgap to separate the occupied and unoccupied bands. Therefore, hybrid functionals are not suited for metallic systems.

\begin{figure}[t!] 
\includegraphics[width=\linewidth]{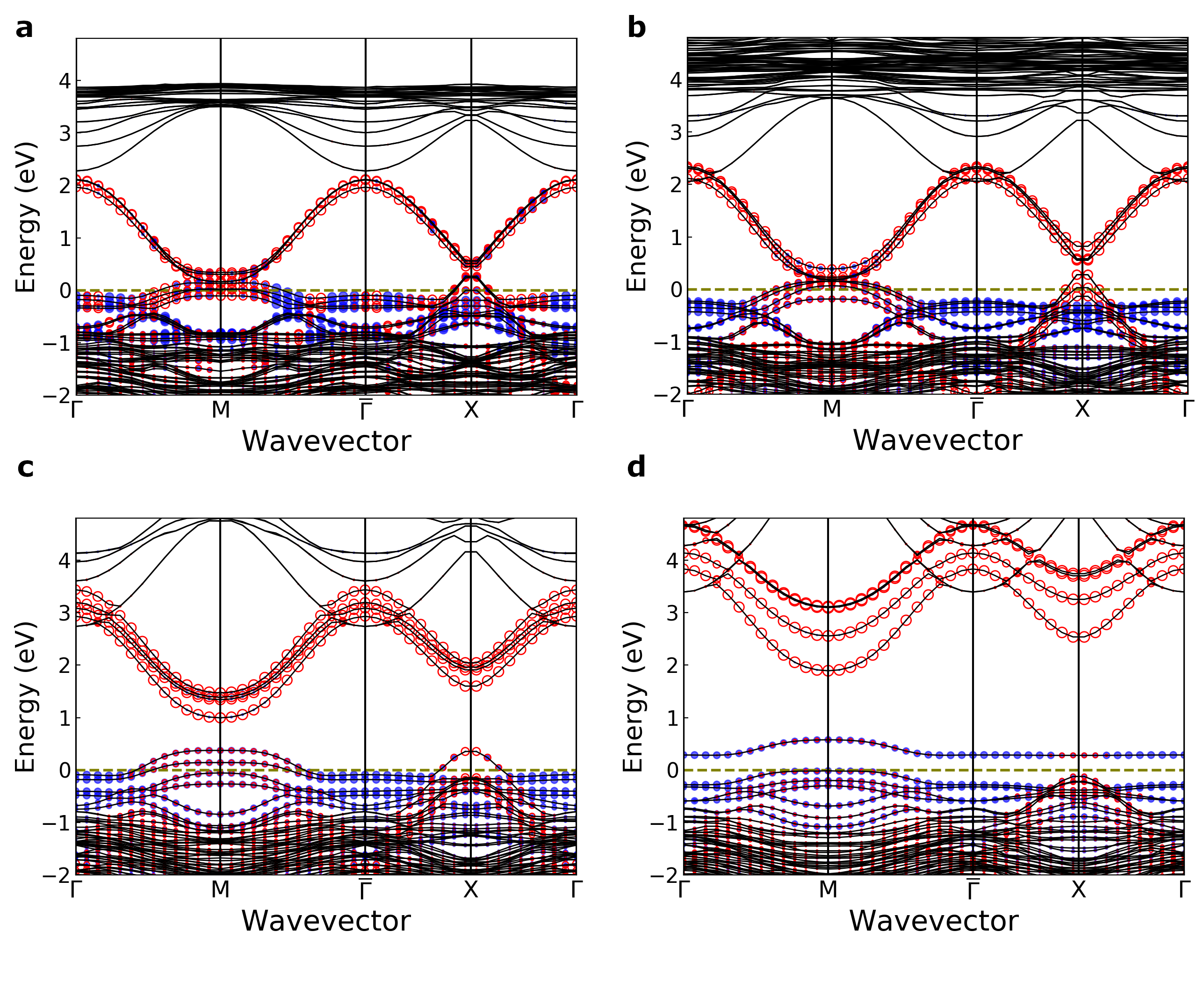} 
\caption{Band structure comparison by varying mixing parameter in the HSE06 hybrid functional for $(\textbf{a})$ \(a\) = 0 $(\textbf{b})$ \(a\) = 0.05 $(\textbf{c})$ \(a\) = 0.15 and $(\textbf{d})$ \(a\) = 0.25 in doped LTO phase. The blue filled and red empty circles correspond to copper d$_\mathrm{z^{2}}$ and copper d$_{\mathrm{x^{2}}-\mathrm{y^{2}}}$  orbitals respectively. The projection strength is denoted by marker size.  } 
\label{fig:aexx}
\end{figure}

With the preceding consideration in mind, it is reasonable that the HSE06 XC functional produces an insulator in LCO but fails to capture the metal-insulator transition under doping. Figure \ref{fig:aexx} shows HSE06 based band structures of LSCO for various mixing parameters \enquote{\(a\)}. For \(a\) = 0, HSE06 is reduced to PBE, and thus predicts LSCO to be metallic. At \(a\) =0.05, a slight change in the band structure can be seen: the conduction bands are slightly pushed up and split due to the stabilization of the magnetic moments on Cu, and the bands around the Fermi level at $X$ start to separate from one another. Increasing \(a\) to 0.15 results in a separation of the valence bands at the Fermi level and the splitting of the conduction bands, and the two valence bands near the Fermi level split off from the remaining valence bands. Finally, at the standard value of \(a\)= 0.25, the highest valence band completely splits off, leaving a 0.2 eV gap at the Fermi level. 
The resulting conduction band displays significant spin-splitting, indicative of a strong uncompensated ferri-magnetic order. Our spin density calculations show that the spin-down band now lies just above the Fermi level where the doped hole is localized in the copper d$_\mathrm{z^{2}}$ and apical-oxygen p$_\mathrm{z}$ hybridized band, see in Figure S5(a) of the Supplementary Material. Moreover, the band-projected charge density for the spin-down band (Supplementary Figure S5(b)) clearly displays a d$_\mathrm{z^{2}}$ orbital shape for copper sites and a p$_\mathrm{z}$ orbital shape on the apical oxygen sites, similar to the results from B3LYP \cite{perry2002ab}.

Band structures of FIG. \ref{fig:aexx} show that for small values of the mixing parameter (\(a\)), the conduction band and valence bands around the M point near the Fermi level are more dominated by copper d$_{\mathrm{x^{2}}-\mathrm{y^{2}}}$ states. As the value of the mixing parameter increases, the copper d$_\mathrm{z^{2}}$ orbitals gain more weight. This implies that, as the fraction of exact exchange increases, electrons are more localized on in-plane copper atoms. This is expected since the LSDA gives the extreme covalent regime while the Hartree-Fock leads to the extreme ionicity.

Our study demonstrates that the meta-GGA class of XC functionals within the generalized Kohn-Sham scheme correctly predict many experimental results for pristine LCO, and also capture the insulator-to-metal transition with Sr doping. Among the different meta-GGAs considered, SCAN’s performance for structural, electronic, and magnetic properties of LCO/LSCO is closest to the corresponding experimental results. In contrast, the hybrid XC functional (HSE06) fails to capture the metal-insulator transition and overestimates the magnetic moments and bandgaps in pristine LCO, and it needs adjustment of the standard 25$\%$ value of the mixing parameter to produce the metallic states. Our study thus indicates that the meta-GGAs provide a robust  new pathway for the  first-principles treatment of strongly correlated materials. 

\section*{METHODS}
\subsection*{Computational methods}
The calculations were performed using the pseudopotential projector-augmented wave method \cite{kresse1999ultrasoft} implemented in the Vienna ab initio simulation package (VASP)\cite{kresse1996efficient,kresse1993ab}. The energy cutoff for the plane-wave basis set was taken to be 550 eV for all meta-GGA calculations and 520 eV for HSE06-based computations. In order to sample the Brillouin zone, for meta-GGAs, an $8\times 8\times4$ $\Gamma$-centered k-point mesh was used while a smaller mesh of $6\times6\times 2$ was used for the HSE06 hybrid functional. The structures were initially relaxed for meta-GGAs using conjugate gradient algorithm with an atomic force tolerance of 0.008 eV/$\ang$ and total energy tolerance of $\mathrm{10 ^{-5}}$ eV. HSE06  calculations on the doped systems were carried out using a damped algorithm. The computational cost for HSE06 based computations is much larger than for the meta-GGAs, and for this reason, a smaller number of k-points and less strict energy tolerance were used in conjunction with using the unrelaxed (experimental) structures. 

\section*{data availability}
The data that supports the findings of this study are available from the corresponding author upon reasonable request.

\pagebreak{}
\bibliographystyle{ieeetr}{}
\bibliography{reference}

\section*{acknowledgments}
The work at Tulane University was supported by the US Department of Energy under EPSCoR Grant No.~DE-SC0012432 with additional support from the Louisiana Board of Regents, by the Donors of the American Chemical Society Petroleum Research Fund, and by the U.S. DOE, Office of Science, Basic Energy Sciences grant number DE-SC0019350. Calculations were partially done using the National Energy Research Scientific Computing Center and the Cypress cluster at Tulane University. The work at Northeastern University was supported by the US Department of Energy (DOE), Office of Science, Basic Energy Sciences grant number DE-FG02-07ER46352, and benefited from Northeastern University's Advanced Scientific Computation Center (ASCC), and the NERSC supercomputing center through DOE grant number DE-AC02-05CH11231. C.L. was supported by the U.S. DOE NNSA under Contract No. 89233218CNA000001 through the LDRD program and by the Center for Integrated Nanotechnologies, a DOE BES user facility, in partnership with the LANL Institutional Computing Program for computational resources. B.B. acknowledges support from the COST Action CA16218.

\section*{author contributions}
J.S. designed the project. K.P. and J.S. proposed the framework of the computational approach, K.P. performed the calculations. C.L., J.W.F.,R.Z., J.N., B.B., R.S.M., Y.Z., A.B. and J.S. analyzed the data, and wrote the manuscript.

\section*{competing interests}
The authors declare no competing interests.

\end{document}


\title{Supplementary Materials for \enquote{Sensitivity of the electronic and magnetic structures of cuprate superconductors to density functional approximations}} 
\author
{Kanun Pokharel,$^{1,*}$  Christopher Lane,$^{2,3,*}$,James W. Furness,$^{1}$Ruiqi Zhang,$^{1}$ Jinliang Ning,$^{1},$ Bernardo Barbiellini,$^{4,5}$ Robert S. Markiewicz,$^{5}$ Yubo Zhang,$^{1}$ Arun Bansil,$^{*5}$ Jianwei Sun$^{*1}$

\textit{\normalsize{$^{1}$Department of Physics and Engineering Physics, Tulane University, New Orleans, LA 70118, USA}}\\
\textit{\normalsize{$^{2}$Theoretical Division, Los Alamos National Laboratory, Los Alamos, New Mexico 87545, USA}}\\
\textit{\normalsize{$^{3}$Center for Integrated Nanotechnologies, Los Alamos National Laboratory, Los Alamos, New Mexico 87545, USA}}\\
\textit{\normalsize{$^{4}$LUT University, P.O. Box 20, FI-53851, Lappeenranta, Finland}}\\
\textit{\normalsize{$^{5}$Physics Department, Northeastern University, Boston, Massachusetts 02115, USA}}\\

\normalsize{$^\ast$To whom correspondence should be addressed; 

E-mail:  kpokhare@tulane.edu, laneca@lanl.gov, ar.bansil@neu.edu and jsun@tulane.edu}
}

\date{\today}
\maketitle
\beginsupplement
\clearpage{}
\clearpage{}

\section*{Contents}
Section S1 Predicted lattice constants for LCO and doped LSCO
\\
Section S2 Octahedral tilt angles for pristine LCO
\\
Section S3 Lattice constant calculations for solids(set SL20)\\
Section S4 Predicted electronic bandgap\\
Section S5 Copper magnetic moment calculations by varying the Wigner-Seitz radius 
\\
Section S6 Electronic structure of the LTO  phase using various XC functionals
\\
Section S7 Electronic structure of the LTT  phase using various XC functionals
\\
Section S8 Electronic structure of the HTT phase using various XC functionals
\\
Section S9 HSE06 PDOS and charge density of the conduction band at $a = 0.25$
\\
Section S10 DOS plot for $a = 0.15$ for the pristine LCO using the HSE06 XC functional
\pagebreak{}
\section*{Section S1: Predicted lattice constants for LCO and doped LSCO}

\begin{table}[h!]
\caption{ Theoretically  predicted values of lattice constants using various meta-GGAs for three different phases of pristine LCO and doped LSCO systems }
\resizebox{\columnwidth}{!}{%
\begin{tabular}{lc|c|cccc|cccc}
\hline
& & &\multicolumn{4}{c|}{ \textbf{pristine LCO}}& \multicolumn{4}{c}{\textbf{doped LSCO}} \\
 \hline
 & Meta-GGA & Phase & \multicolumn{4}{c|}{Lattice constant} & \multicolumn{4}{c}{Lattice constant}  \\ 
&  & &a($\ang$) & b($\ang$) & c($\ang$) & V($\ang^3$) & a($\ang$) & b($\ang$) & c($\ang$) & V($\ang^3$) \\
 \hline\hline
 & & LTO&5.335 &5.421 &13.107 &379.1&- & -& -&- \\
 & {Experimental} & LTT  &5.360 &5.360 &13.236 &380.3&- & -& -&- \\
 & & HTT &5.391&5.391 &13.219 &384.2&- &- &- & -\\
 \hline\hline
 & & LTO&5.328 &5.494 &13.064 &382.54&5.329 & 5.429& 13.090&378.74 \\
 & {TPSS} & LTT  &5.417 &5.417 &13.059 &383.3&5.374 &5.376 &13.097 &378.5 \\
 & & HTT &5.370&5.370 &13.118&378.37&5.355 & 5.351& 13.118& 375.99\\
\hline\hline
 & & LTO&5.328 &5.473 &13.043 &380.42&5.326 & 5.416& 13.062&376.84 \\
 & {revTPSS} & LTT  &5.401 &5.401 &13.046 &380.62&5.365 &5.370 &13.071 &376.63 \\
 & & HTT &5.366&5.366 &13.089&376.92&5.349 & 5.349& 13.085& 374.39\\
 \hline\hline
 & & LTO&5.376 &5.415&13.241&385.58&5.356&5.365 &13.271 & 381.42\\
 & {M06L} & LTT &5.380&5.380&13.246&383.53 &5.358 &5.361 &13.287 & 381.79\\
 & & HTT &5.385&5.385&13.250&384.28 &5.376 &5.376 &13.270 &383.58 \\
 \hline\hline
 & & LTO&5.33&5.50&13.040&382.86&5.333 &5.452 &13.052 &379.57 \\
 & {MS0} & LTT &5.420&5.420&13.048&383.41&5.362 &5.374 &13.171 &379.59 \\
 & & HTT &5.368&5.368&13.080&377&5.359 &5.358 &13.061 &375.1 \\
 \hline\hline
 & & LTO&5.312&5.47&13.018&378.34 &5.314 &5.400 &13.047 &374.49 \\
 & {MS2} & LTT &5.3914&5.391&13.025&378.66&5.357 &5.371 &13.043 &375.33 \\
 & & HTT &5.349&5.349&13.055&373.6&5.331 &5.329 &13.073 &371.48 \\
 \hline\hline
 & & LTO&5.323&5.455&13.086&380.07&5.321 &5.402 &13.077 &375.95 \\
 & {SCAN} & LTT & 5.391&5.391&13.080&380.21&5.353 &5.364 &13.090 &375.96 \\
 & & HTT &5.349&5.349&13.125&375.6 &5.335 &5.333 &13.106 &372.98\\
 \hline\hline
 & & LTO&5.317&5.421&13.18&380.12&5.294 &5.368 &13.228 &375.98 \\
 & {SCAN-L} & LTT &5.368&5.367&13.177&379.68 &5.332 &5.330 &13.228 &376 \\
 & & HTT & 5.341&5.341&13.21&378.26&5.315 &5.312 &13.241& 373.92 \\
 \hline\hline
 & & LTO&5.321&5.460&13.088&380.32&5.31 &5.385 &13.126 &375.43 \\
 & {rSCAN} & LTT &5.388&5.388&13.092&380.14 &5.345 &5.356 &13.118 &375.64 \\
 & & HTT & 5.343&5.343&13.154&375.63&5.328 &5.327 &13.141& 372.99 \\
 \hline\hline
 & & LTO&5.327&5.467&13.099&381.54&5.322 &5.394 &13.131 &376.99 \\
 & {$\mathrm{r^{2}}$SCAN} & LTT &5.394&5.394&13.107&381.38 &5.355 &5.359 &13.131 &376.92 \\
 & & HTT & 5.350&5.350&13.168&376.96&5.338 &5.336 &13.148& 374.59 \\
 \hline
 \end{tabular}
 }
 \label{table:doped}
\end{table}

\clearpage{}

\section* {Section S2: Octahedral tilt angles for pristine LCO}

\begin{table}[H]
\centering
\caption{Theoretically  predicted values of octahedra tilt using various meta-GGAs for three different phases of parent LCO system  }
\begin{tabular}{lc|c|c}
 \hline 
  & Meta-GGA & Phase & Octahedral Tilt \\
 & & &axial (deg) \\ 
 \hline\hline 
 & & LTO&6.930  \\
 & {TPSS} & LTT &6.880 \\
 & & HTT &0  \\
 \hline\hline
 & & LTO&6.460  \\
 & {revTPSS} & LTT &6.360 \\
 & & HTT &0  \\
 \hline\hline
 & & LTO& 3.760 \\
 & {M06L} & LTT&3.270  \\
 & & HTT &0 \\
 \hline\hline
 & & LTO &8.320  \\
 & {MS0} & LTT& 8.280\\
 & & HTT& 0 \\
 \hline\hline
 & & LTO&7.320 \\
 & {MS2} & LTT& 7.350 \\
 & & HTT&0  \\
 \hline\hline
 & & LTO&7.090  \\
 & {SCAN} & LTT &7.010  \\
 & & HTT &0 \\
 \hline\hline
 & & LTO&5.740  \\
 & {SCAN-L} & LTT& 5.470  \\
 & & HTT& 0  \\
 \hline\hline
 & & LTO&6.930  \\
 & {rSCAN} & LTT& 6.630  \\
 & & HTT& 0  \\
 \hline\hline
 & & LTO&6.910  \\
 & {$\mathrm{r^{2}}$SCAN} & LTT& 6.500  \\
 & & HTT& 0  \\
 \hline
 \end{tabular}
 \label{tab:pristine}
 \end{table}
\pagebreak{} 

\section* {Section S3: Lattice constant calculations for solids (set SL20)}

We performed lattice constant calculations for solids (set SL20) using energy-volume (E-V) curve approach and compared the results against those obtained from a full relaxation of the structures using the $\mathrm{r^{2}}$SCAN-L XC functional. This comparison showed a significant difference in the predicted lattice constants for many systems given in Table \ref{table:lattice}. Additionally, the lattice constants from E-V curve  are closer to reference values than the fully relaxed lattice constants, suggesting that there is an error in the $\mathrm{r^{2}}$SCAN-L stress tensor implementation. To further confirm the error, we performed the calculations also by employing thee $\mathrm{r^{2}}$SCAN functional,see Table \ref{table:lattice}. The lattice constants obtained via the two  approaches are close and comparable to the reference values. For this reason, we have not included the lattice constants and octahedral tilt values for $\mathrm{r^{2}}$SCAN-L in the main text. Note that the lattice constant calculation for the cuprate using E-V curve could not carried out since the crystal structure involves a large number of degrees of freedom. For the SL20 set, lattice constants are identical (a=b=c) while in the cuprate they are not, which makes the calculations more difficult.  
\begin{table}[h!]
\centering
\caption{SL20 calculations using full relaxation and E-V curve approaches are compared for $\mathrm{r}^{2}$SCAN-L and $\mathrm{r}^{2}$SCAN functional approximations. The mean absolute deviation(MAD) between two approaches against the reference values are also listed. }
\begin{tabular}{lc|ccc|ccc }
 &&&$\mathrm{r}^{2}$SCAN&& & $\mathrm{r}^{2}$SCAN-L& \\
 & {Solids} & Reference[1] & {EV}&{Relaxation}&{Reference[2]}&{EV}&{Relaxation}\\
 \hline
 &Ag&4.096&4.100&4.107&4.138&4.139&4.029\\
 \hline
 &Al&3.986&3.987&3.988&3.973&3.971&3.801\\
 \hline
 &C&3.558&3.560&3.559&3.571&3.570&3.566\\
 \hline
 &Ca&5.574&5.572&5.570&5.503&5.500&5.480\\
 \hline
 &Cu&3.575&3.578&3.581&3.609&3.602&3.480\\
 \hline
 &GaAs&5.669&5.670&5.671&5.694&5.690&5.620\\
 \hline
 &Ge&5.683&5.682&5.680&5.705&5.700&5.594\\
 \hline
 &Li&3.460&3.459&3.466&3.430&3.435&3.568\\
 \hline
 &LiF&3.993&3.994&3.993&4.005&4.000&3.915\\
 \hline
 &Na&4.218&4.210&4.221&4.139&4.182&4.299\\
 \hline
 &NaF&4.610&4.598&4.589&4.511&4.594&4.505\\
 \hline
 &Rh&3.802&3.801&3.802&3.824&3.827&3.813\\
 \hline
 &SiC&4.352&4.353&4.355&4.356&4.354&4.321\\
 \hline
 &LiCl&5.104&5.106&5.109&5.073&5.086&5.119\\
 \hline
 &MgO&4.216&4.214&4.210&4.197&4.208&4.179\\
 \hline
 &NaCl&5.605&5.601&5.600&5.509&5.563&5.621\\
 \hline
 &Pd&3.908&3.908&3.910&3.938&3.942&3.916\\
 \hline
 &Si&5.439&5.439&5.436&5.427&5.426&5.345\\
 \hline
 &Sr&6.096&6.098&6.100&6.045&6.048&6.348\\
 \hline
 &Ba&5.078&5.071&5.074&5.051&5.051&5.335\\
 \hline
 &MAD&0&0.003&0.004&0&0.013&0.097\\
 
 \end{tabular}
 \label{table:lattice}
\end{table}
\pagebreak{}

\clearpage{}

\section* {Section S4: Predicted electronic bandgap}
\begin{table}[h!]
\centering
\caption{Theoretically predicted values of electronic bandgap by using various XC functionals }
\begin{tabular}{lccc }
 \hline
 & {Functionals} & {Phase} & {electronic bandgap (eV)} \\ 
\\
\hline\hline
 & & LTO&0.180\\
 & {TPSS} & LTT &0.250\\
 & & HTT &0.210\\
 \hline\hline
 & & LTO&0.210\\
 & {revTPSS} & LTT&0.270\\
 & & HTT &0.240\\
 \hline\hline
 & & LTO&0.540\\
 & {M06L} & LTT &0.520\\
 & & HTT &0.540\\
 \hline\hline
 & & LTO&1.220 \\
 & {MS0} & LTT&1.270\\
 & & HTT &1.180\\
 \hline\hline
 & & LTO&1.020\\
 & {MS2} & LTT &1.050\\
 & & HTT &0.980\\
 \hline\hline
 & & LTO&1.090\\
 & {SCAN} & LTT &1.120\\
 & & HTT &1.030\\
 \hline\hline
 & & LTO&0.430\\
 & {SCAN-L} & LTT &0.470\\
 & & HTT &0.460\\
 \hline\hline
 & & LTO&0.790\\
 & {rSCAN} & LTT &0.840\\
 & & HTT &0.780\\
 \hline\hline
 & & LTO&0.840\\
 & {$\mathrm{r^{2}}$SCAN} & LTT &0.870\\
 & & HTT & 0.820\\
 \hline\hline
 & & LTO&0.470\\
 & {$\mathrm{r^{2}}$SCAN-L} & LTT &0.530\\
 & & HTT &0.490\\
 \hline\hline
 & & LTO&3.200\\
 & {HSE06} & LTT &3.100\\
 & & HTT &3.000\\
\hline
 \end{tabular}
 \label{table:mag_ind}
\end{table}
\pagebreak{}

\clearpage{}

\section* {Section S5: Copper magnetic moments calculated by varying the Wigner-Seitz radius}
\begin{figure}[h!] 
\begin{center}
\includegraphics[width=0.7\linewidth]{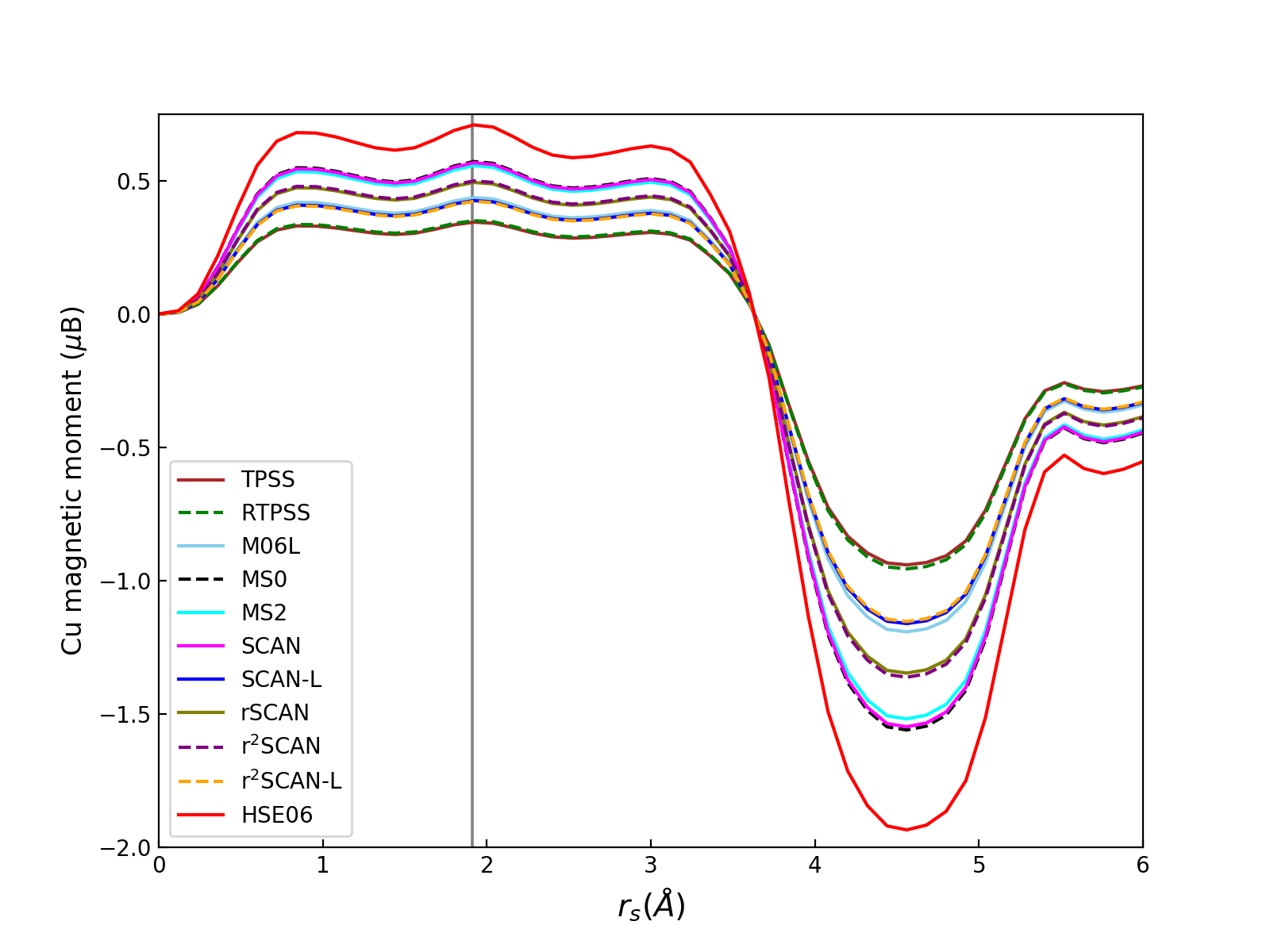} 
\caption{Copper magnetic moments of $\mathrm{La_{2}CuO_{4}}$ in the LTO crystal structure by various functional approximations when Wigner-Seitz radius is varied. The vertical grey line represents Cu-O bond length.}
\label{fig:rwigs}
\end{center}
\end{figure}

We computed the copper (Cu) magnetic moment by varying the Wigner-Seitz radius ($r_{s}$) beyond the default of 1.16 \AA~ to the value of 1.91 \AA~ (the Cu-O bond length) to fully capture the magnetic density centered on the Cu site and the part of the magnetic density originating from the strong hybridization between the copper and oxygen atoms. Cu magnetic moments as a function of the Wigner-Seitz radius based on using various functionals are shown in Figure \ref{fig:rwigs}: slightly enhanced magnetic moments are seen for all functionals when the overlap between the Cu and oxygen atoms is taken into account. At $r_{s}$ of 1.91 \AA, we fully enclose the Cu site, resulting in the maximum positive magnetic moment $(S_0)$ for all functional approximations. As $r_{s}$ increases further, the negative density from the tails of the first nearest-neighbor shell of 4 atoms start to contribute. At $r_{s}$ $\sim$ 3.63 \AA, we approximately capture 1/4 of each nearest-neighbor site, yielding a moment of zero due to the staggered magnetization. For $r_{s}$ $\sim$ 4.5 \AA, the Wigner-Seitz sphere now fully encompasses the central atomic site and the first nearest-neighbor shell, producing a large negative value since $S=S_0+4S_{nn}$, where $S_{nn}$ is the magnetic moment value on the nearest-neighbor sites and $S_{nn}=-S_0$. Note that the Cu moment for large $r_s$ values will depend on the coordination number and thus on the crystal structure. Moments for the three phases considered based on various functionals are shown in Table \ref{table:mag}.

\begin{table}[h!]
\centering
\caption{Theoretically predicted values of copper magnetic moments by various XC functionals }
\begin{tabular}{lccc }
 \hline
 & {Functionals} & {Phase} & {Cu magnetic moments ($\mu_{B}$)} \\ 
\\
\hline\hline
 & & LTO& 0.340\\
 & {TPSS} & LTT & 0.340\\
 & & HTT &0.310\\
 \hline\hline
 & & LTO&0.350\\
 & {revTPSS} & LTT&0.350\\
 & & HTT &0.340 \\
 \hline\hline
 & & LTO& 0.430\\
 & {M06L} & LTT &0.430 \\
 & & HTT &0.440\\
 \hline\hline
 & & LTO&0.580 \\
 & {MS0} & LTT&0.580\\
 & & HTT &0.560\\
 \hline\hline
 & & LTO& 0.550\\
 & {MS2} & LTT&0.550 \\
 & & HTT &0.540\\
 \hline\hline
 & & LTO&0.560\\
 & {SCAN} & LTT &0.560\\
 & & HTT &0.550\\
 \hline\hline
 & & LTO&0.420\\
 & {SCAN-L} & LTT &0.420\\
 & & HTT &0.420\\
 \hline\hline
 & & LTO&0.490\\
 & {rSCAN} & LTT &0.490\\
 & & HTT &0.470\\
 \hline\hline
 & & LTO&0.490\\
 & {$\mathrm{r^{2}}$SCAN} & LTT &0.490\\
 & & HTT &0.470\\
 \hline\hline
 & & LTO&0.420\\
 & {$\mathrm{r^{2}}$SCAN-L} & LTT&0.420\\
 & & HTT &0.410\\
 \hline\hline
 & & LTO&0.700\\
 & {HSE06} & LTT &0.690\\
 & & HTT &0.690\\
\hline
 \end{tabular}
 \label{table:mag}
\end{table}
\pagebreak{}

\clearpage{}

\section*{Section S6. Electronic structure of the LTO phase using various XC functionals}
\begin{figure}[h!] 
\begin{center}
\includegraphics[width=0.67\linewidth]{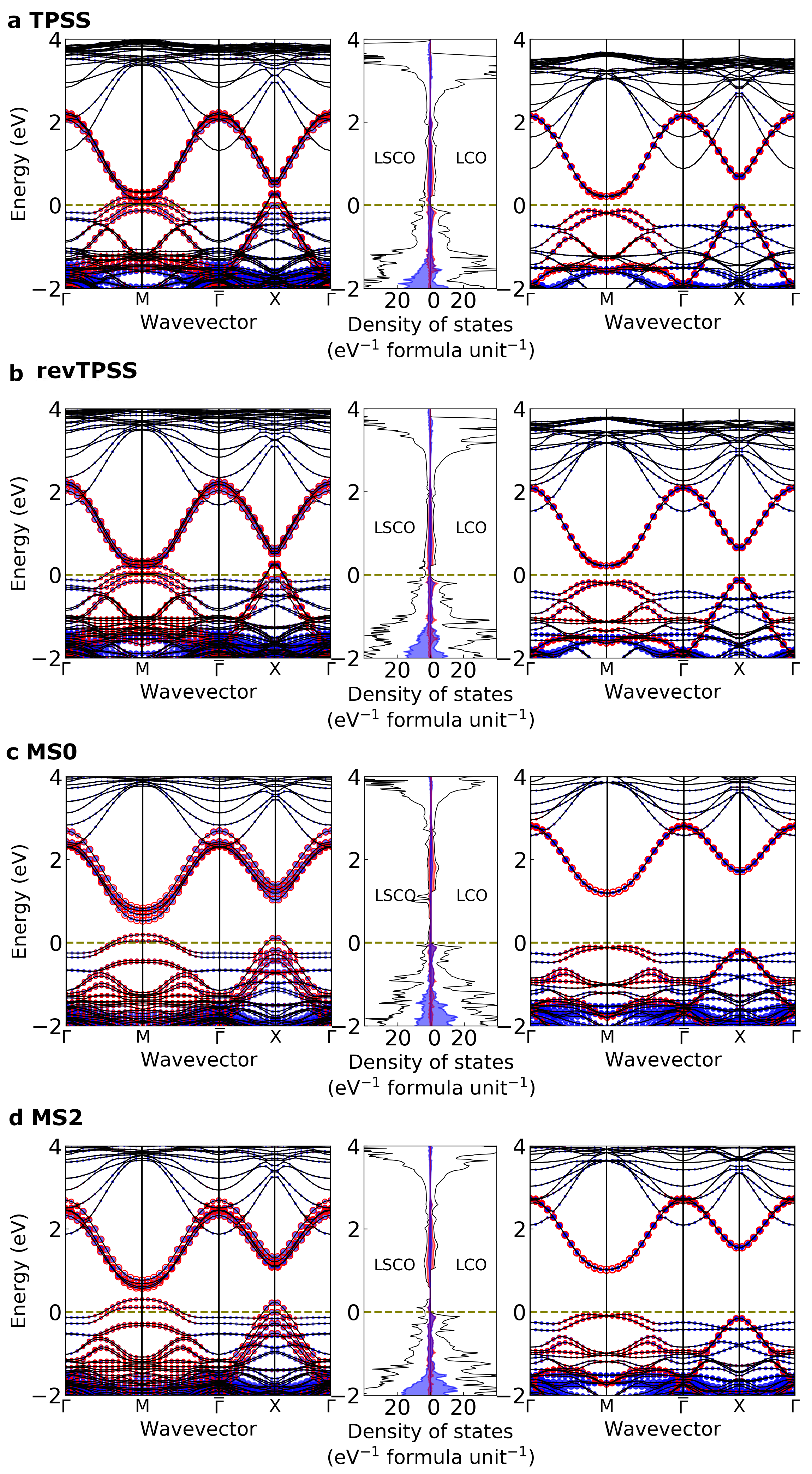} 
\label{fig:lto_1}
\end{center}
\end{figure}
\begin{figure}[h!] 
\begin{center}
\includegraphics[width=0.75\linewidth]{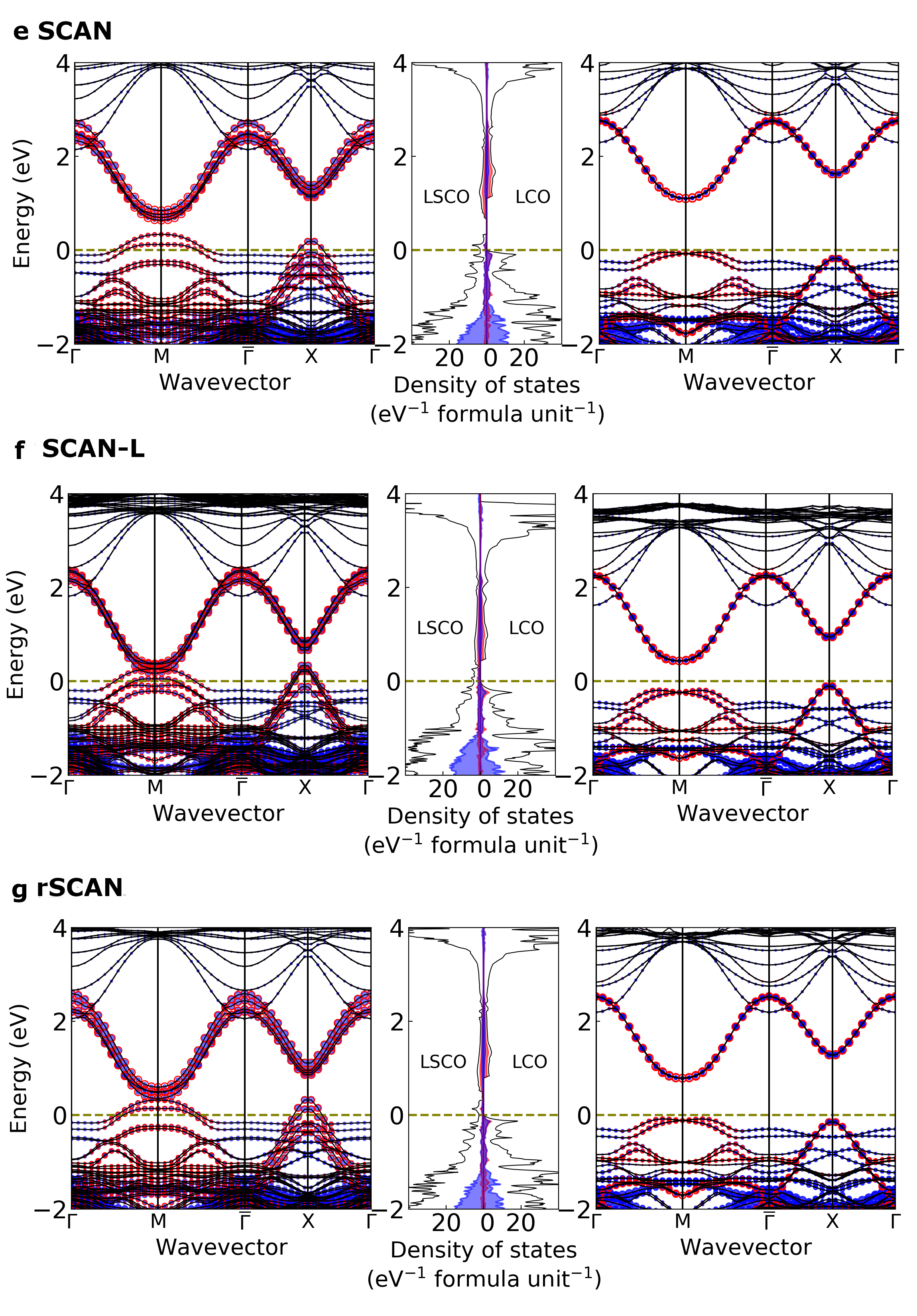} 
\caption{Electronic structure results for  $(\textbf{a}) $ TPSS $,(\textbf{b}) $ revTPSS $,(\textbf{c}) $ MSO $,(\textbf{d}) $ MS2 $,(\textbf{e}) $ SCAN $,(\textbf{f})$ SCAN-L, and $(\textbf{g}) $ rSCAN for pristine LCO and doped LSCO systems for the LTO phase.}
\label{fig:lto_2}
\end{center}
\end{figure}
\clearpage{}

\section*{Section S7. Electronic structure of the LTT phase by various XC functionals}
\begin{figure}[h!]
\begin{center}
\includegraphics[width=0.75\linewidth]{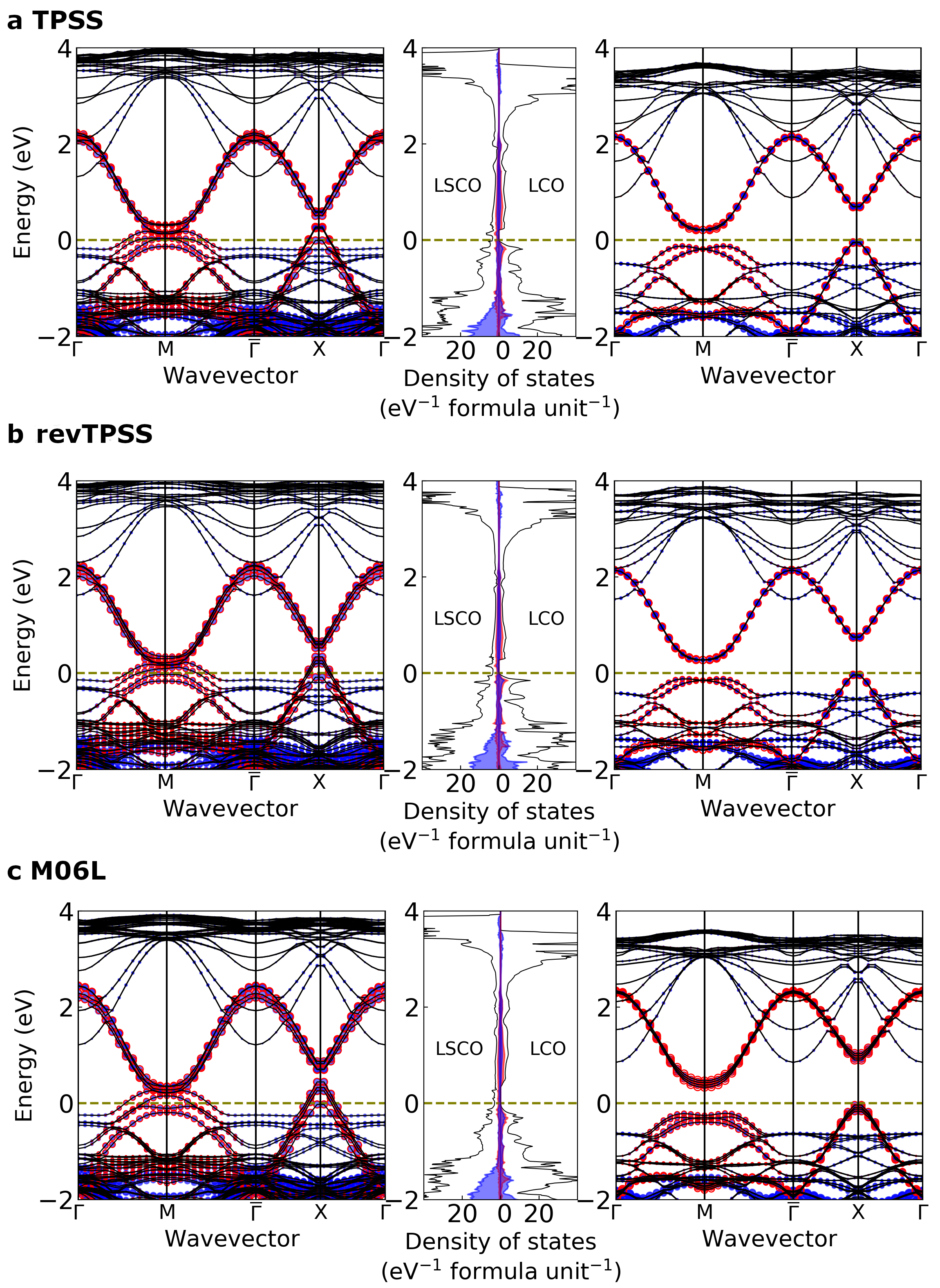} 
\label{fig:ltt_1}
\end{center}
\end{figure}
\begin{figure}[h!] 
\begin{center}
\includegraphics[width=0.8\linewidth]{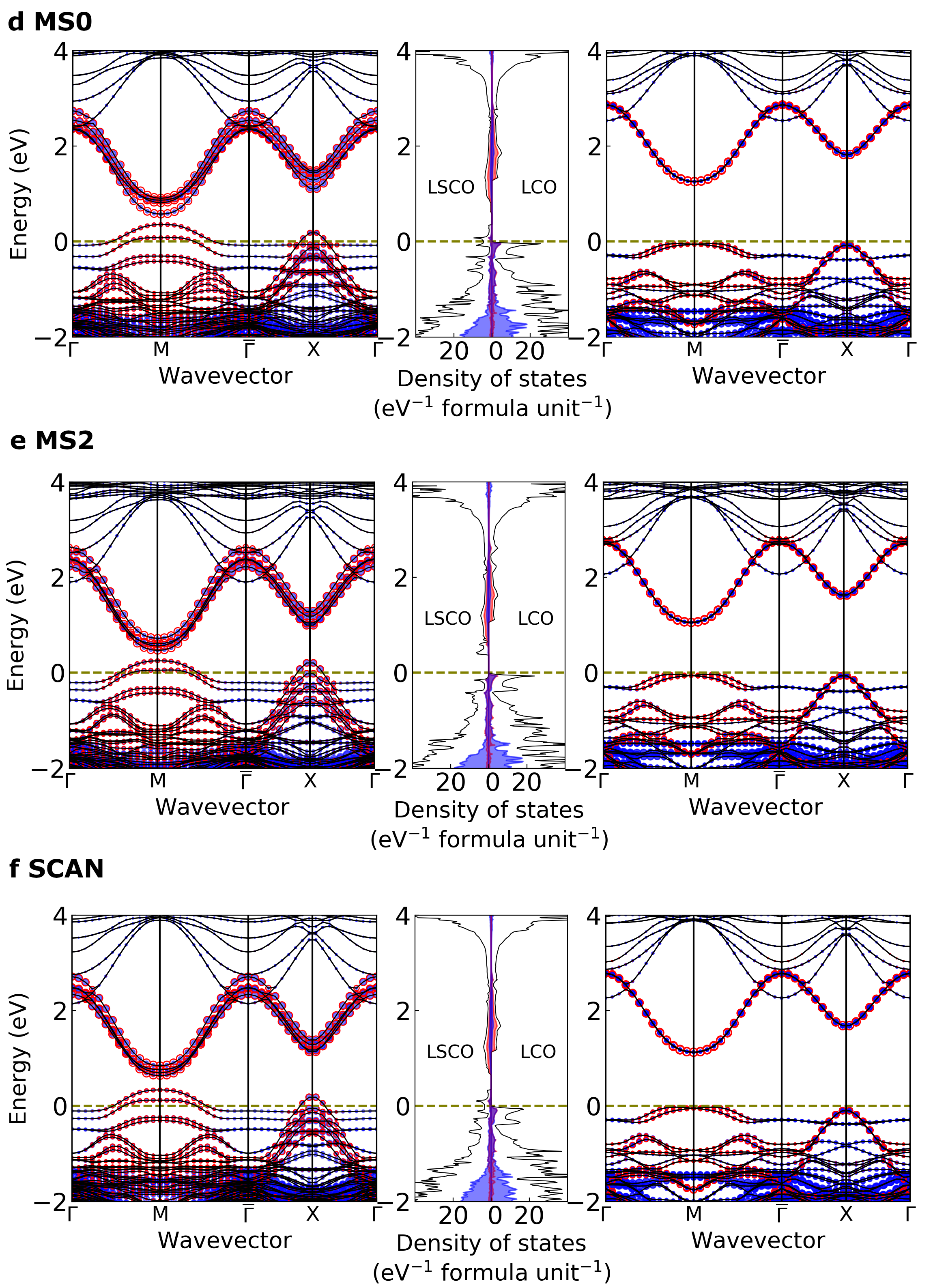} 
\label{fig:ltt_2}
\end{center}
\end{figure}
\begin{figure}[h!] 
\begin{center}
\includegraphics[width=0.69\linewidth]{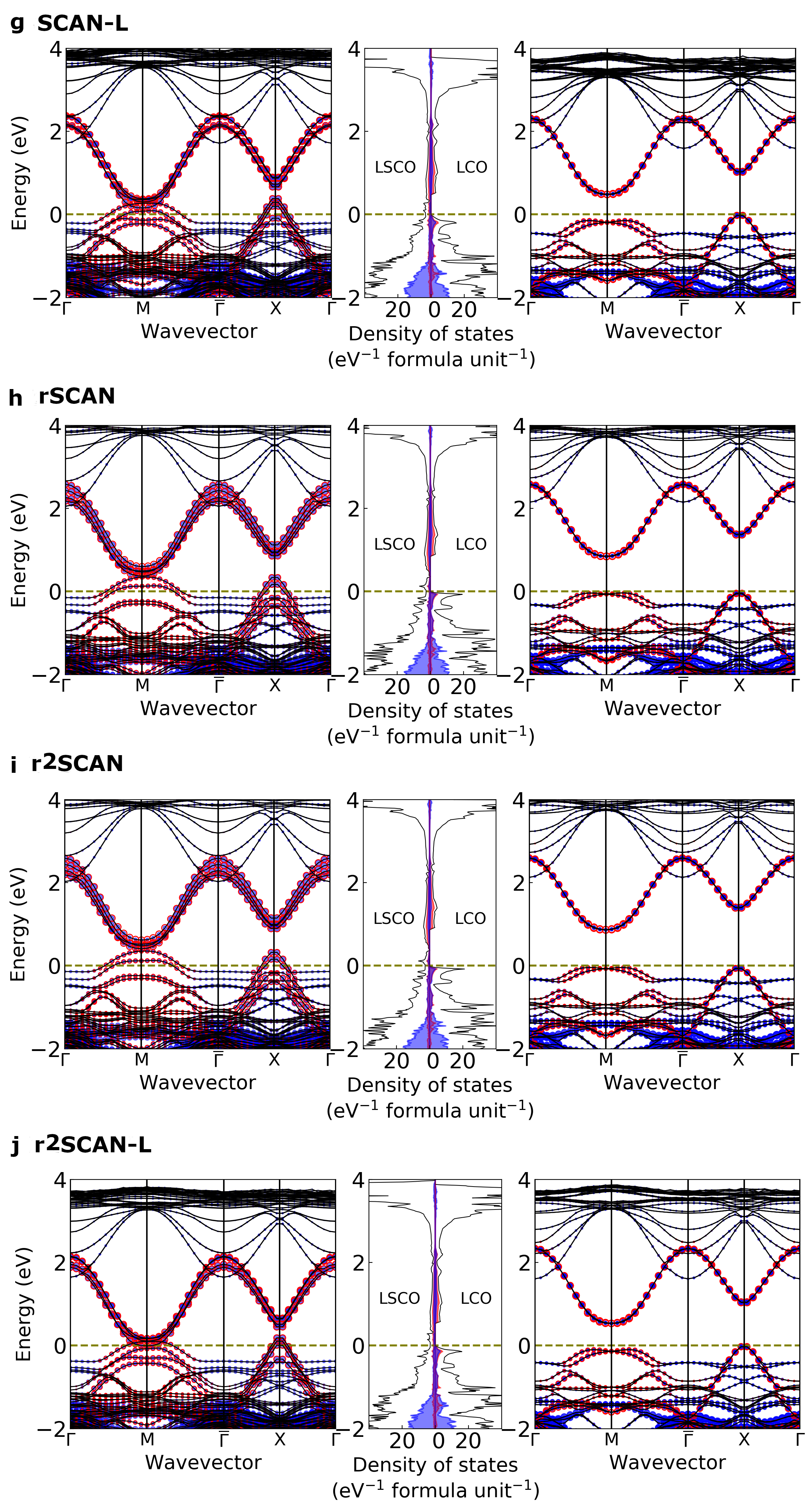} 
\caption{Electronic structure results for  $(\textbf{a}) $ TPSS $,(\textbf{b}) $ revTPSS, $(\textbf{c}) $ M06L $,(\textbf{d}) $ MS0 $,(\textbf{e}) $ MS2 $,(\textbf{f})$ SCAN $,(\textbf{g}) $ SCAN-L $,(\textbf{h}) $ rSCAN $,(\textbf{i}) $ $\mathrm{r^{2}}$SCAN, and $(\textbf{j}) $ $\mathrm{r^{2}}$SCAN-L for pristine LCO and doped LSCO systems for the LTT phase. See Ref.[3] for the LTT crystal structure used.}
\label{fig:ltt_3}
\end{center}
\end{figure}

\clearpage{}

\section* {Section S8. Electronic structure of the HTT phase using various XC functionals}
\begin{figure}[h!] 
\begin{center}
\includegraphics[width=0.8\linewidth]{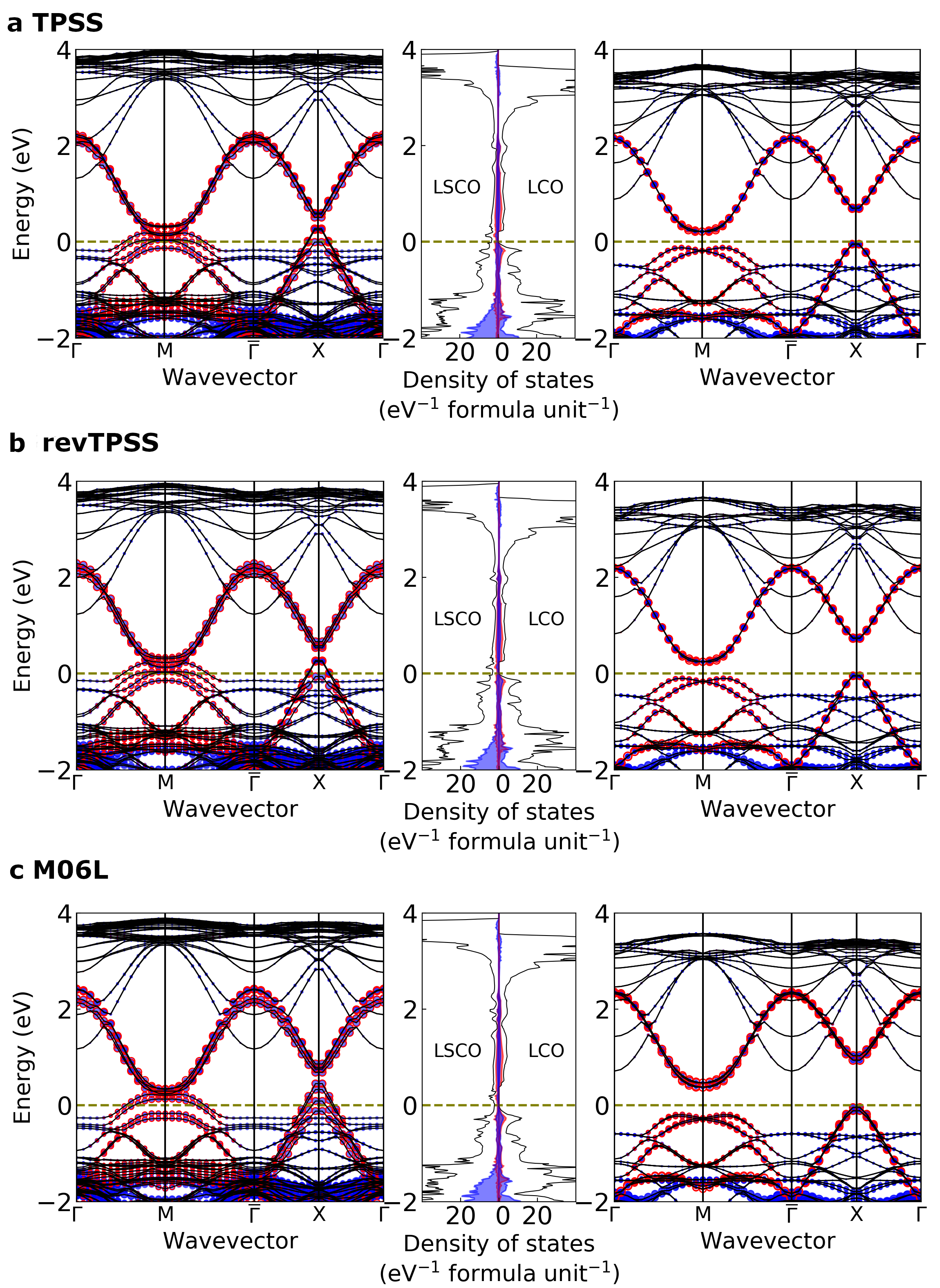} 
\label{fig:htt_1}
\end{center}
\end{figure}

\begin{figure}[h!] 
\begin{center}
\includegraphics[width=0.8\linewidth]{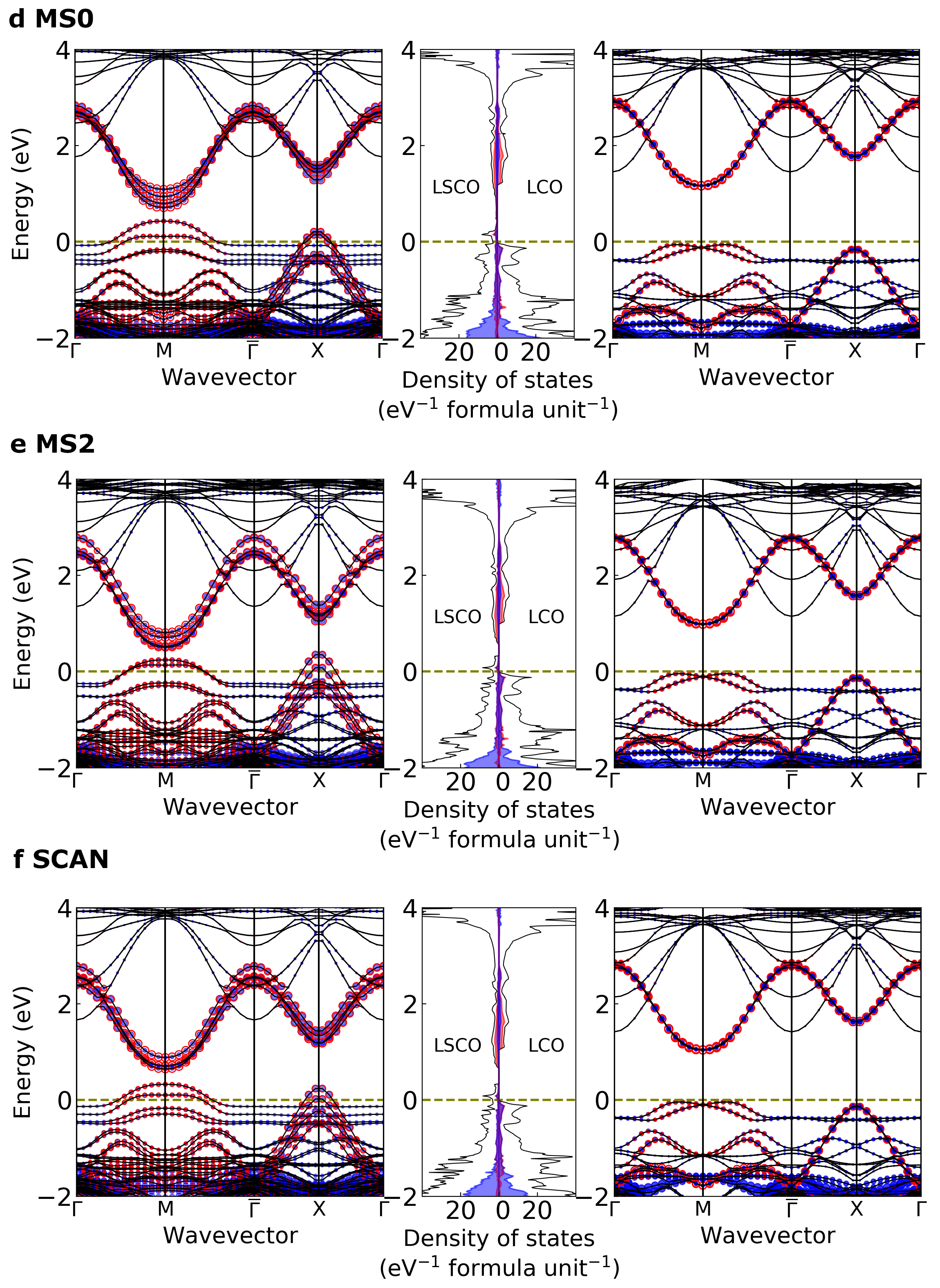} 
\label{fig:htt_2}
\end{center}
\end{figure}

\begin{figure}[h!] 
\begin{center}
\includegraphics[width=0.7\linewidth]{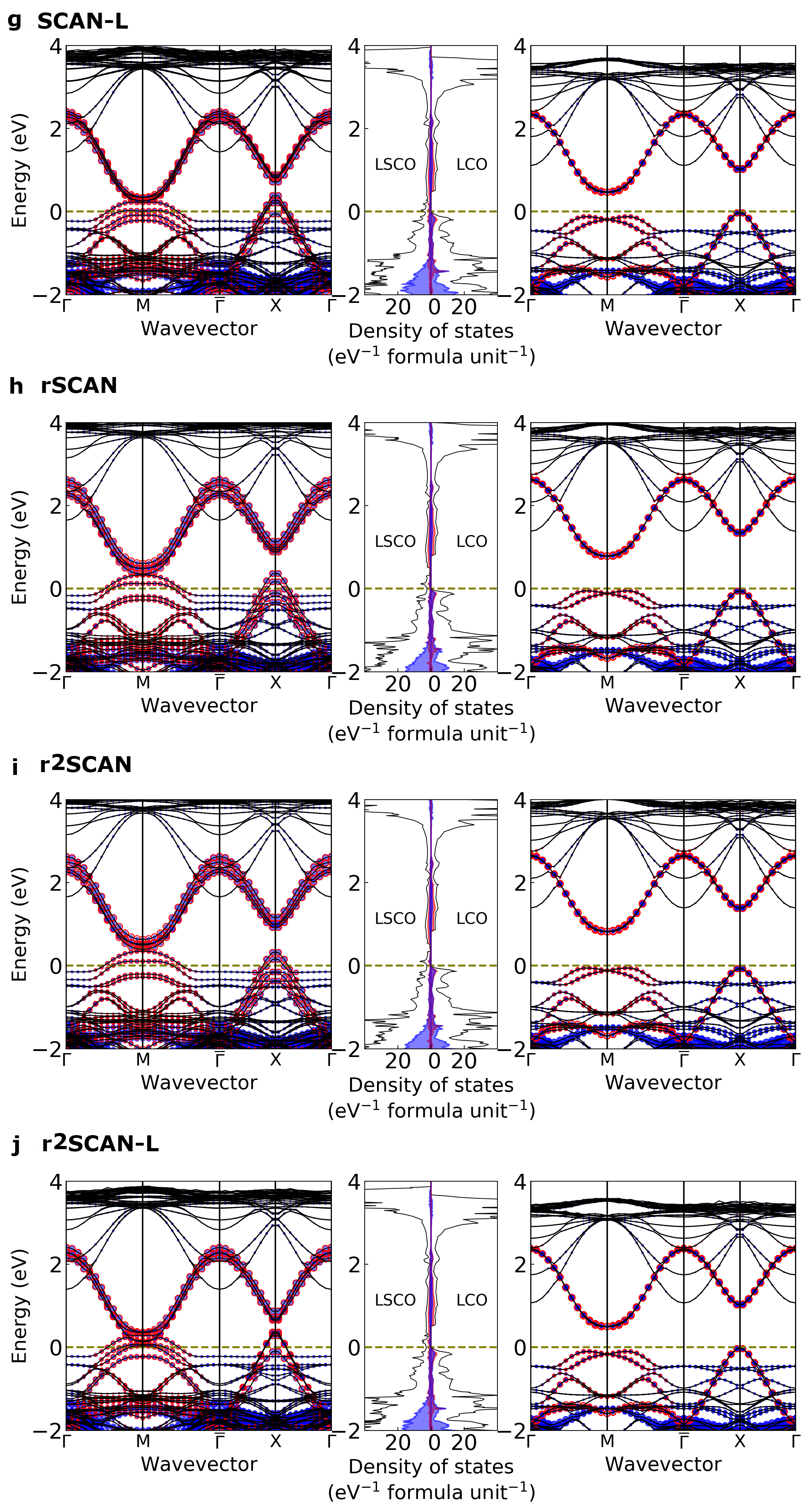} 
\caption{Electronic structure results for  $(\textbf{a}) $ TPSS $,(\textbf{b}) $ revTPSS $,(\textbf{c}) $ M06L $,(\textbf{d}) $ MS0 $,(\textbf{e}) $ MS2 $,(\textbf{f})$ SCAN $,(\textbf{g}) $ SCAN-L $,(\textbf{h}) $ rSCAN $,(\textbf{i}) $ $\mathrm{r^{2}}$SCAN $(\textbf{j}) $, and $\mathrm{r^{2}}$SCAN-L for pristine LCO and doped LSCO systems for HTT phase. See Ref.[3] for the HTT crystal structure used.}
\label{fig:htt_3}
\end{center}
\end{figure}
\clearpage{}

\section* {Section S9: HSE06 PDOS and charge density of the conduction band at \(a\) = 0.25} 
\begin{figure}[h] 
\begin{center}
\includegraphics[width=1\linewidth]{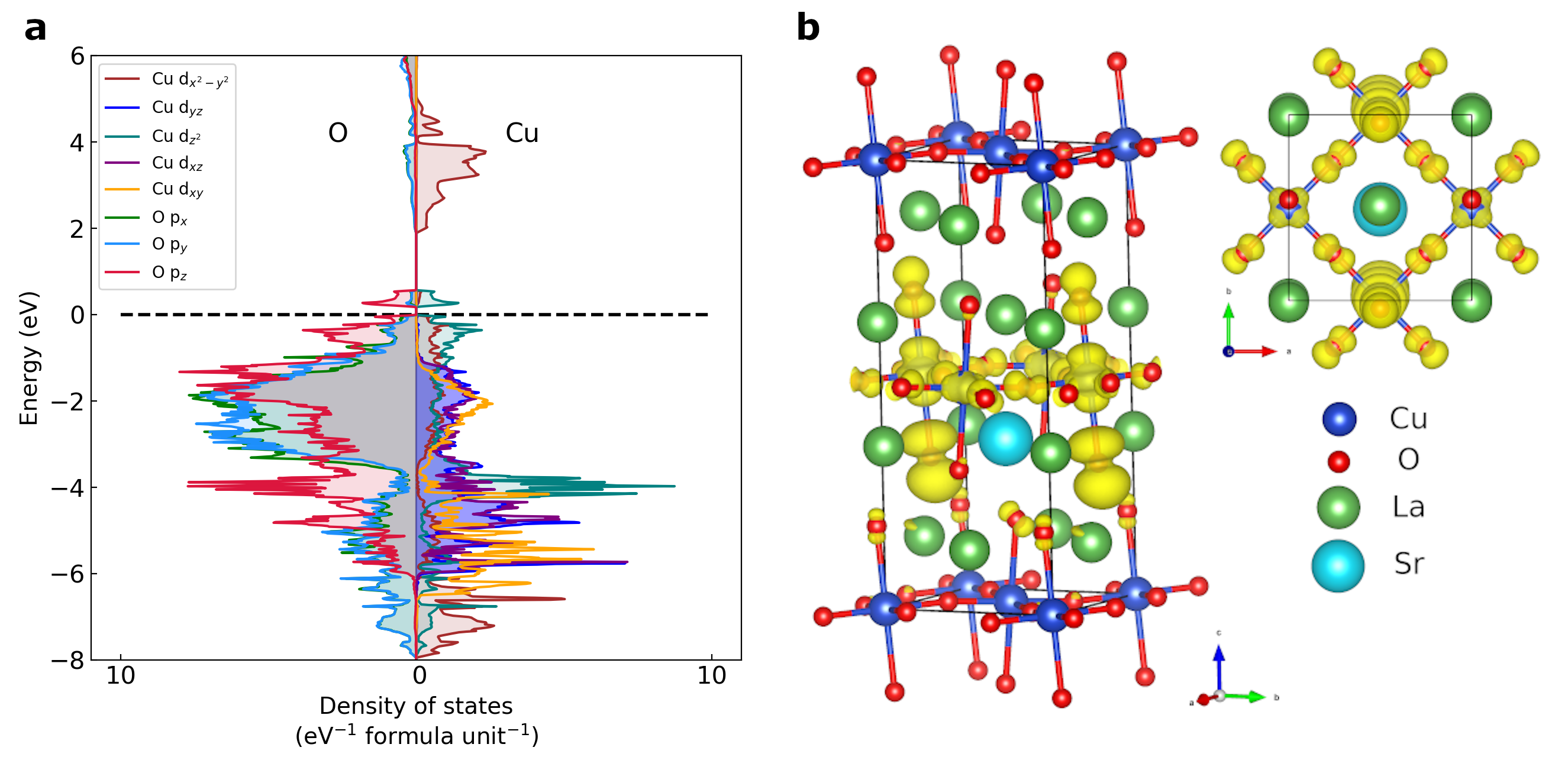} 
\caption{$(\textbf{a})$ represents PDOS for copper atom on the right side and oxygen atom on the left side of the plot. $(\textbf{b})$ represents spin density plot for the conduction band at \(a\) = 0.25. The doped hole is found to be localized within copper and oxygen atoms shown by yellow iso-surface which is $\mathrm{d_{z^2}}$ for Cu and $\mathrm{p_{z}}$ for O in nature.} 
\label{fig:spind}
\end{center}
\end{figure}

\clearpage{}

\section*{Section S10: DOS plot for \(a\) = 0.15 for the pristine LCO using the HSE06 XC functional}
\begin{figure}[h] 
\begin{center}
\includegraphics[width=0.95\linewidth]{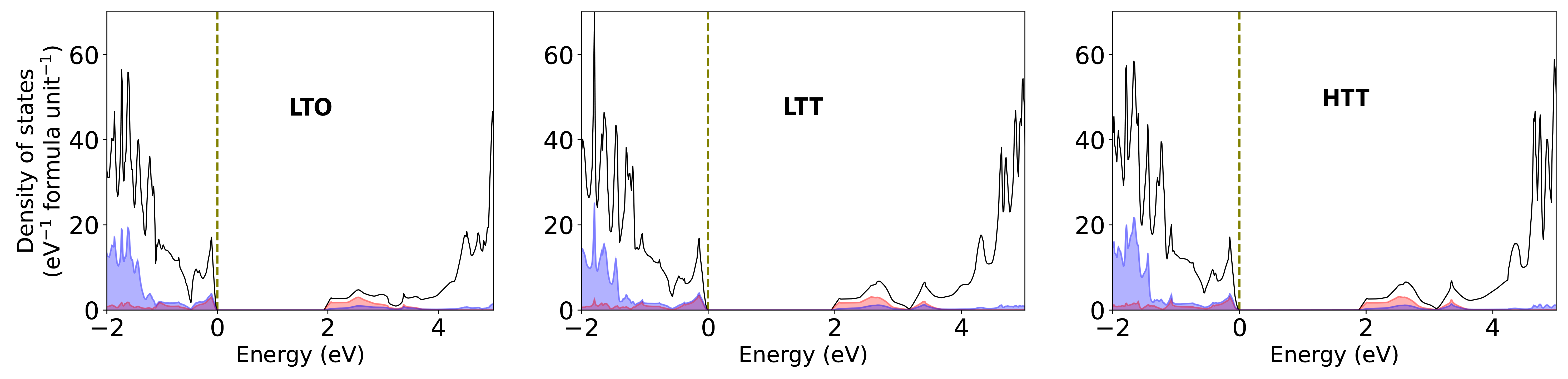} 
\caption{DOS plot of LTO, LTT and HTT for the value of \(a\) = 0.15 for pristine LCO using HSE06 functional approximation. The plot shows a reduced band gap and reduced magnetic moment, similar to the results for the doped system.   }
\label{fig:hse_0.15}
\end{center}
\end{figure}

\clearpage{}
\section*{References}
$\hspace{1ex}\hspace{1ex}\hspace{1.5ex}$[1] J. W. Furness, A. D. Kaplan, J. Ning, J. P. Perdew, and J. Sun, “Accurate and
Numerically Efficient r2SCAN Meta-Generalized Gradient Approximation,” J. Phys.
Chem. Lett., vol. 11, no. 19, pp. 8208–8215, 2020.

[2] D. Mej$\acute{i}$a- Rodr$\acute{i}$guez and S. Trickey, “Meta-GGA performance in solids at almost GGA
cost,” Phys. Rev. B, vol. 102, no. 12, p. 121109, 2020.

[3] J. W. Furness, Y. Zhang, C. Lane, I. G. Buda, B. Barbiellini, R. S. Markiewicz, A. Bansil, and J. Sun, “An accurate first-principles treatment of doping-dependent electronic
structure of high-temperature cuprate superconductors,” Commun. Phys., vol. 1, no. 1,
p. 11, 2018